# Analytical description of the size effect on pyroelectric and electrocaloric properties of ferroelectric nanoparticles

Anna N. Morozovska [1,2,*], Eugene A. Eliseev[3], Maya D. Glinchuk[3], Hanna V. Shevliakova[4], George S. Svechnikov[4], Maxim V. Silibin [5,6,7], Artem V. Sysa[8], Andrii D. Yaremkevich[1], Nicholas V. Morozovsky[1], and Vladimir V. Shvartsman[9,†]

[1] *Institute of Physics, National Academy of Sciences of Ukraine,*
*46, pr. Nauky, 03028 Kyiv, Ukraine*

[2] *Bogolyubov Institute for Theoretical Physics, National Academy of Sciences of Ukraine,*
*14-b Metrolohichna str. 03680 Kyiv, Ukraine*

[3] *Institute for Problems of Materials Science, National Academy of Sciences of Ukraine,*
*Krjijanovskogo 3, 03142 Kyiv, Ukraine*

[4] *Department of Microelectronics, National Technical University of Ukraine "Igor Sikorsky Kyiv Polytechnic Institute", Kyiv, Ukraine*

[5] *National Research University of Electronic Technology, 124498 Zelenograd, Moscow, Russia*

[6] *Institute for Bionic Technologies and Engineering, I. M. Sechenov First Moscow State Medical University, 119991 Moscow, Russia*

[7] *Scientific-Practical Materials Research Centre of NAS of Belarus, 220072 Minsk, Belarus*

[8] *Research and Manufacturing Complex Technology Center "MIET",*
*124498 Zelenograd, Moscow, Russia*

[9] *Institute for Materials Science and Center for Nanointegration Duisburg-Essen (CENIDE),*
*University of Duisburg-Essen, 45141, Essen, Germany*

**Abstract**

Using Landau-Ginzburg-Devonshire theory and effective medium approximation, we analytically calculate typical dependences of the pyroelectric and electrocaloric coefficients on external electric field, temperature and radius for spherical single-domain ferroelectric nanoparticles. The considered physical model corresponds to the nanocomposite with small fraction of ferroelectric nanoparticles. Within the framework of the analytical model we establish how the size changes determine the temperature and field behavior pyroelectric and electrocaloric coefficients on example of $BaTiO_3$ nanoparticles covered by a semiconducting shell and placed in a dielectric polymer. We show that by changing the particle size one can induce maxima of the pyroelectric coefficient and electrocaloric temperature variation, control their width, height and sign. Obtained analytical expressions allow selecting the interval of particle sizes, voltage, and/or temperature for which pyroelectric energy conversion and electrocaloric coefficient are optimal for applications. The observed size effect opens the possibility to control pyroelectric and electrocaloric properties of ferroelectric nanocomposites that can be important for their advanced applications in energy convertors and cooling systems.

---

[*] corresponding author, e-mail: anna.n.morozovska@gmail.com

[†] corresponding author, e-mail: vladimir.shvartsman@uni-due.de

## I. INTRODUCTION

Nanosized ferroelectrics attract permanent attention of researchers as unique model objects for fundamental studies of polar surface properties, various screening mechanisms of spontaneous polarization by free carriers, and possible emergence of versatile multi-domain states [1, 2, 3, 4, 5, 6]. This fully applies to ferroelectric nanoparticles, for which effective procedures of synthesis and methods of polar properties control have been developed. Classical examples are experimental results of Yadlovker and Berger [7, 8], who revealed the ferroelectricity enhancement in Rochelle salt cylindrical nanoparticles. Frey and Payne [9], Zhao et al. [10], Drobnich et al. [11], Erdem et al. [12], Shen et al. [13], and Golovina et al. [14] demonstrated the possibility to control phase transition temperatures and other features of $BaTiO_3$, $Sn_2P_2S_6$, $PbTiO_3$, $SrBi_2Ta_2O_9$, and $KTa_{1-x}Nb_xO_3$ nanopowders and nanoceramics by finite size effects.

The continuum phenomenological Landau-Ginzburg-Devonshire (**LGD**) approach combined with the electrostatic equations allows establishing the physical origins of the anomalies in the polar and dielectric properties of ferroelectric nanoparticles and predicts changes of their phase diagrams when the particle size decreases. For instance, using the LGD approach Perriat et. al. [15], Huang et al. [16], Glinchuk et al. [17], Ma [18], Khist et al. [19], Wang et al. [20, 21], Eliseev et al. [22, 23] and Morozovska et al. [24, 25, 26], showed that the transition temperatures, the degree of spontaneous polar ordering in spherical, ellipsoidal, and cylindrical ferroelectric nanoparticles of size of 4 – 100 nm are conditioned by various physical mechanisms, such as surface tension, correlation effect, depolarization field originated from the incomplete screening of spontaneous polarization, flexoelectricity, electrostriction and Vegard-type chemical pressure.

Electrocaloric (**ECE**) and pyroelectric (**PEE**) effects that are inherent to ferroelectrics are the subjects of intensive experimental and theoretical studies [27, 28, 29]. Electrocaloric (**EC**) and pyroelectric (**PE**) properties of ferroelectrics at ferroelectric-antiferroelectric phase boundaries [30], ferroelectric thin films [31, 32, 33, 34], multilayers [35, 36, 37] and other low-dimensional materials [38] can be very different from those of single crystals [39].

As it is known [40, 41], the polar materials in adiabatic conditions are characterized by the PEE (charge or electric field generation under temperature change) and by the inverse ECE (temperature change under application or removal of an electric field). The vivid manifestation of PEE and ECE in ferroelectrics is a consequence of the strong temperature dependence of the spontaneous polarization [42, 43, 44], especially in the vicinity of phase transitions [45, 46] or near the morphotropic phase boundary [47]. This property is the basis for the widespread applications of ferroelectric materials for pyroelectric detectors and energy converters, as well as for realizing their potentiality in modern electrocaloric converters [48, 49, 50].

At present, ECE and PEE in ferroelectric crystals, ceramics and polymers, thin films and multilayer structures are the objects of intensive theoretical, experimental, and applied studies. Nevertheless, ECE and PEE in ferroelectric nanoparticles are relatively poor studied. The possible reason is the strong influence of size effects via depolarization field [24] and polarization-strain coupling [20, 25] on the polarization distribution, ferroelectric transition temperature, dielectric, PE, and EC properties. There are several studies directed on the elucidation of the features of the PEE and ECE in nanowires, nanotubes [51, 52, 53, 54], and nanoparticles [55]. However, the **analytical description** of ECE and PEE in the most "technological" spherical nanoparticles and nanocomposites, allowing for depolarization and incomplete screening effect, is still missing.

Using the LGD theory and effective medium approximation, this work analyzes typical dependences of the polarization, dielectric permittivity, PE and EC coefficients on external electric field, temperature, and radius for spherical ferroelectric nanoparticles covered by a semiconducting shell and placed in a dielectric medium. The considered physical model corresponds to a nanocomposite "nanoparticles-matrix" with a small fraction (less than 10%) of the ferroelectric nanoparticles.

The manuscript has the following structure. Problem statement containing free energy and basic equations with boundary conditions is formulated in **Section II**. **Section III** introduces approximate analytical expressions for the transition temperature, EC temperature change, heat capacity, and related physical quantities. Size effect on ECE and PEE is analyzed in **Section VI** using the example of BaTiO$_3$ nanoparticles**. Section V** presents analysis of the size effect on the PE and EC energy conversion. **Section VI** contains conclusive remarks. Calculation details of the transition temperature, PEE and ECE, and auxiliary figures are presented in **Appendixes A, B, C** and **D,** respectively.

## II. PROBLEM STATEMENT

Let us consider a spherical ferroelectric nanoparticle of radius *R* covered by a semiconducting shell of thickness Λ and placed in a dielectric medium (polymer, gas, liquid, air or vacuum) with an effective isotropic dielectric permittivity $\varepsilon_e$ [**Fig. 1**].

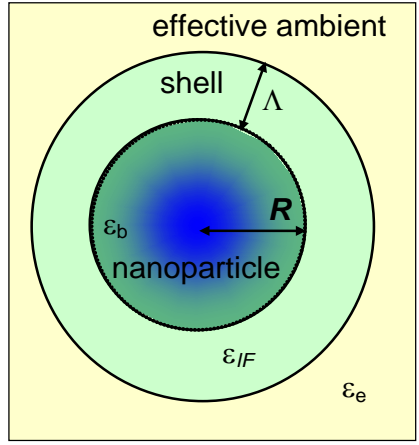

**FIGURE 1.** A spherical ferroelectric nanoparticle (core) covered by a semiconducting layer (shell) and placed in a paraelectric or dielectric ambient.

A nanoparticle in a ferroelectric phase has a one-component spontaneous polarization $P_3(\mathbf{r})$ directed along the crystallographic axis 3. The dependence of other electric polarization components on the inner electric field $E_i$ is linear, $P_i = \varepsilon_0(\varepsilon_b - 1)E_i$, where $i = 1, 2$, $\varepsilon_b$ is an isotropic relative permittivity of background [56], and $\varepsilon_0$ is the universal dielectric constant. Since the ferroelectric polarization component $P_3(\mathbf{r})$ contains both background and soft mode contributions, electric displacement vector has the form $\mathbf{D} = \varepsilon_0\varepsilon_b\mathbf{E} + \mathbf{P}$ inside the particle. Outside the particle $\mathbf{D} = \varepsilon_0\varepsilon_e\mathbf{E}$. The electric field components $E_i$ are related with the electric potential φ as $E_i = -\partial\varphi/\partial x_i$. The potential φ satisfies Poisson equation inside the particle and Laplace equation outside it:

$$\varepsilon_0\varepsilon_b\left(\frac{\partial^2}{\partial x_1^2} + \frac{\partial^2}{\partial x_2^2} + \frac{\partial^2}{\partial x_3^2}\right)\varphi = \begin{cases} \dfrac{\partial P_3}{\partial x_3}, & r < R, \\ 0, & r > R \end{cases} \qquad (1a)$$

Equations (1) are supplemented by the conditions of potential continuity at the particle surface, $\left(\varphi_{ext} - \varphi_{int}\right)\big|_{r=R} = 0$ and field homogeneity at infinity, $\varphi_{ext}\big|_{r\to\infty} = -x_3 E_{ext}$. Here $E_{ext}$ is the external electric field far from the particle (which can be absent). The boundary condition for the normal components of the electric displacements is $\left(\mathbf{n}(\mathbf{D}_{ext} - \mathbf{D}_{int}) + \sigma_S\right)\big|_{r=R} = 0$. The "effective" surface charge density $\sigma_S$ is introduced to model realistic conditions of the spontaneous polarization

incompletely screened at the ferroelectric particle surface and depends on its ambient (dielectric, inert or chemically active gases, liquids, semiconductor or imperfect electrode cover). Several theoretical studies [19, 23, 26] use the linear dependence of the charge density $\sigma_S$ on the electric potential excess at the surface of the nanoparticle, $\delta\varphi = \varphi_{int}|_{r=R} - \varphi_{ext}|_{r\to\infty}$:

$$\sigma_S[\varphi] \approx -\varepsilon_0 \varepsilon_{IF} \frac{\delta\varphi}{\Lambda}, \qquad (1b),$$

where an "effective" screening length $\Lambda$ [26] and the interfacial dielectric permittivity $\varepsilon_{IF}$ are introduced.

It should be noted that the expression (1b) is approximate because it includes an "effective" screening charge, while the real space charge is distributed in a ultrathin layer near the interface [57] or imperfect electrodes with nonzero screening length [58]. Stengel et al. [59, 60] introduced the concept of the interfacial capacitance $C_{IF}$ for the $\Lambda$ description. Actually, $C_{IF} = \varepsilon_0 \varepsilon_{IF} S / \Lambda$ (in a flat capacitor approximation) allows to justify the Eq.(1b), because the product $C_{IF} \varphi|_{r=R}$ is the total value of the interfacial space charge, $q = \sigma_S S$, and therefore $\sigma_S = \dfrac{C_{IF} \varphi|_{r=R}}{S} \approx -\varepsilon_0 \varepsilon_{IF} \dfrac{\varphi|_{r=R}}{\Lambda}$.

To fulfill the inequality $R \gg \Lambda$, reliable estimations of the $\Lambda$ value should be used. Following Wang et al. [61] and Tagantsev et al. [62], the effective screening length $\Lambda$ (more rigorously, $\Lambda/\varepsilon_{IF}$), should be much smaller than 1 Å (about 0.1Å) in accordance with modern *ab initio* estimations [62, 63]. One of the reasons why the "effective" $\Lambda/\varepsilon_{IF}$ can be much smaller than the typical perovskite lattice constant $a \sim 0.5$ nm [62], is the high relative dielectric permittivity $\varepsilon_{IF}$ in the double electric layer, which typically is more than 100.

Another important case (relevant to the nanoparticles suspension in chemically active gases or liquids) is the Stephenson-Highland (**SH**) ionic adsorption at the ferroelectric surface [64, 65]. Within SH model the dependence of the surface charge density $\sigma_S[\varphi]$ on the electric potential excess $\delta\varphi$ at the free surface is controlled by the concentration of positive and negative surface charges in a self-consistent manner via Langmuir adsorption isotherms as is shown in Refs.[6, 26].

Since we would not like to be limited to a specific model, further we perform calculations for $\Lambda$ changing in the range (0.1 – 10) nm, and $\varepsilon_{IF} > 100$ to provide an effective screening of the nanoparticle spontaneous polarization.

LGD free energy functional $G$ additively includes 2-4-6 Landau expansion on polarization powers, $G_{Landau}$, polarization gradient energy contribution, $G_{grad}$, electrostatic contribution $G_{el}$, elastic, electrostriction, and flexoelectric contributions $G_{es+flexo}$. Following Ref.[26] it has the form:

$$G = G_{Landau} + G_{grad} + G_{el} + G_{es+flexo}, \qquad (2a)$$

$$G_{Landau} = \int_{|\vec{r}|<R} d^3r \left( \frac{\alpha}{2} P_3^2 + \frac{\beta}{4} P_3^4 + \frac{\gamma}{6} P_3^6 \right), \tag{2b}$$

$$G_{grad} = \int_{|\vec{r}|<R} d^3r \left( \frac{g_{11}}{2} \left( \frac{\partial P_3}{\partial x_3} \right)^2 + \frac{g_{44}}{2} \left[ \left( \frac{\partial P_3}{\partial x_2} \right)^2 + \left( \frac{\partial P_3}{\partial x_1} \right)^2 \right] \right), \tag{2c}$$

$$G_{el} = -\int_{|\vec{r}|<R} d^3r \left( P_3 E_3 + \frac{\varepsilon_0 \varepsilon_b}{2} E_i E_i \right) - \int_{|\vec{r}|=R} d^2r \frac{\sigma_S \varphi}{2} - \frac{\varepsilon_0 \varepsilon_e}{2} \int_{|\vec{r}|>R} E_i E_i d^3r, \tag{2d}$$

$$G_{es+flexo} = \int_{|\vec{r}|<R} d^3r \left( -\frac{s_{ijkl}}{2} \sigma_{ij} \sigma_{kl} - Q_{ij3} \sigma_{ij} P_3^2 - F_{ijk3} \left( \sigma_{ij} \frac{\partial P_3}{\partial x_k} - P_3 \frac{\partial \sigma_{ij}}{\partial x_k} \right) \right). \tag{2e}$$

The coefficient $\alpha$ linearly depends on temperature $T$, $\alpha = \alpha_T(T - T_C)$, where $T_C$ is the Curie temperature and $\alpha_T$ is the inverse Curie-Weiss constant. Coefficients $\beta$ and $\gamma$ can be temperature-dependent (e.g. for BaTiO$_3$) but in another way, $\beta = \beta_T(T - T_\beta)$ and $\gamma = \gamma_T(T - T_\gamma)$. Coefficient $\beta$ is positive in the case of a 2$^{nd}$ order ferroelectric phase transition (**FEPT**) of and is negative in the case of a 1$^{st}$ order FEPT. The gradient coefficients $g_{11}$ and $g_{44}$ are positive and regarded as temperature independent. In Eq.(2e), $\sigma_{ij}$ is the stress tensor.

We omit the explicit form of the $G_{es+flexo}$ for simplicity; it is described in Refs.[66, 67, 68]. Since the values of the electrostriction and flexoelectric tensor components, $Q_{ijkl}$ and $F_{ijkl}$ respectively, are unknown for many ferroelectrics, we performed numerical calculations with the coefficients varied in a physically reasonable range ($|F_{ijkl}| \leq 10^{11}$ m$^3$/C, $|Q_{ijkl}| \leq 0.1$ m$^4$/C$^2$). Numerical results for BaTiO$_3$ proved the insignificant impact of electrostriction and flexoelectric coupling.

Allowing for Khalatnikov mechanism of polarization relaxation, minimization of the free energy (2) with respect to the polarization $P(\mathbf{r}_3)$ leads to the time-dependent LGD-equation [26]:

$$\Gamma \frac{\partial P_3}{\partial t} + \alpha_T(T - T_C^*)P_3 + \beta P_3^3 + \gamma P_3^5 - g_{44}\left(\frac{\partial^2}{\partial x_1^2} + \frac{\partial^2}{\partial x_2^2}\right)P_3 - g_{11}\frac{\partial^2 P_3}{\partial x_3^2} = E_3. \tag{3a}$$

The Khalatnikov kinetic coefficient $\Gamma$ determines the relaxation time of the polarization $\tau_K = \Gamma/|\alpha|$ that typically varies in the range $10^{-11} - 10^{-13}$ s far from $T_C$. The boundary condition for the polarization at the spherical surface $r = R$ is natural, $\partial \vec{P}_3/\partial \mathbf{n}\big|_{r=R} = 0$, $\mathbf{n}$ is the outer normal to the surface. Below we also suppose that the external field is $E_{ext} = E_0 \sin(\omega t)$.

The dynamic dielectric susceptibility defined as $\chi_{33} = \dfrac{\partial P_3}{\partial E_3}$ obeys the equation:

$$\Gamma \frac{\partial \chi_{33}}{\partial t} + \left[\alpha_T(T - T_C^*) + 3\beta P_3^2 + 5\gamma P_3^4\right]\chi_{33} - g_{44}\left(\frac{\partial^2}{\partial x_1^2} + \frac{\partial^2}{\partial x_2^2}\right)\chi_{33} - g_{11}\frac{\partial^2 \chi_{33}}{\partial x_3^2} = 1. \tag{3b}$$

The dynamic PE coefficient defined as $\Pi_3 = -\left(\frac{\partial P_3}{\partial T}\right)_E$ obeys the equation:

$$\Gamma \frac{\partial \Pi_3}{\partial t} + \left[\alpha_T(T - T_C^*) + 3\beta P_3^2 + 5\gamma P_3^4\right]\Pi_3 - g_{44}\left(\frac{\partial^2}{\partial x_1^2} + \frac{\partial^2}{\partial x_2^2}\right)\Pi_3 - g_{11}\frac{\partial^2 \Pi_3}{\partial x_3^2} = \alpha_T P_3. \quad (3b)$$

The EC temperature change $\Delta T_{EC}$, can be calculated from the expression:

$$\Delta T_{EC} = -T\int_{E_1}^{E_2} \frac{1}{\rho C_P}\left(\frac{\partial P}{\partial T}\right)_E dE \cong T\int_{E_1}^{E_2} \frac{1}{\rho C_P}\Pi_3 dE, \quad (4a)$$

where $\rho$ is the density, $T$ is the ambient temperature, and $C_p$ is the specific heat. For ferroics the specific heat depends on polarization (and so on external field) and can be modeled as following [69]:

$$C_P = C_P^0 - T\frac{\partial^2 g}{\partial T^2}, \quad (4b)$$

where $C_P^0$ is the polarization-independent part of specific heat and $g$ is the density of the LGD free energy (2). According to experiment, the specific heat usually has a jump at the 2nd order FEPT and has a maximum at the 1st order FEPT, which height is about 10 – 30 % of the $C_p$ value near $T_C$ (see e.g. [35, 70]). Corresponding entropy change is given by expression, $\Delta S = -\int_{E_1}^{E_2}\left(\frac{\partial P}{\partial T}\right)_E dE$.

Finite element modeling (**FEM**) has been performed to find the solution of a coupled equations system (1)-(4) for BaTiO$_3$ nanoparticles placed in a polymer matrix, since such nanocomposites have been intensively studied for energy storage, PEE, and ECE application [71, 72, 73]. We have chosen BaTiO$_3$ because it is a classical proper ferroelectric with the relatively high spontaneous polarization at room temperature, relatively low FEPT temperature and well-known material parameters. BaTiO$_3$ undergoes the 1st order FEPT from the ferroelectric to paraelectric phase. The 1st order FEPT adds additional interesting peculiarities of PE and EC properties, analyzed below, in comparison with the ferroelectric materials undergoing the 2nd order FEPT. Material parameters of BaTiO$_3$ were collected from Refs.[74, 75, 76, 77] and references therein; they are listed in **Table I** and **Table AI, Appendix A.**

**Table I.** LGD parameters for bulk ferroelectric BaTiO$_3$

| $\varepsilon_b$ | $\alpha_T$ (C$^{-2}$·m J/K) | $T_C$ (K) | $\beta$ (C$^{-4}$·m$^5$J) | $\gamma$ (C$^{-6}$·m$^9$J) | $g_{11}$ (m$^3$/F) | $g_{44}$ (m$^3$/F) |
|---|---|---|---|---|---|---|
| 7 | 6.68×10$^5$ | 381 | $\beta_T(T-393)-8.08\times10^8$ <br> $\beta_T = 18.76\times10^6$ | $\gamma_T(T-393)+16.56\times10^9$ <br> $\gamma_T = -33.12\times10^7$ * | 5.1×10$^{-10}$ | 0.2×10$^{-10}$ |

*These parameters are valid until $\gamma > 0$, i.e. for T < 445 K.

** $\rho = 6.02\times10^3$ kg/m$^3$, $C_p = 4.6\times10^2$ J/(kg·K) at room temperature.

FEM results are shown in **Fig. 2**, where we have taken into consideration the multiaxiality of BaTiO$_3$ and use the free energy expansion with the material parameters from **Table AI**. Two sets of initial conditions were used for the polarization distributions, namely bi-domain and single-domain structures. The initial bi-domain structure was transformed into a vortex-like structure with the polarization rotating to align parallel to the particle surface in order to minimize the depolarization electric field. As the result, the latter could not be reduced to zero, but its amplitude becomes significantly smaller (compare the scale in **Fig. 2a** with **Fig. 2b**).

It turned out that the vortex-like domain structure (as electric toroidal multipole) is much less sensitive to homogeneous external field than the single-domain state (electric dipole), unless the field reaches much higher values. So the polarization vortex presents a little interest for PE applications, since it is electro-neutral as a whole. For EC applications, where the polarization response to small external fields should be as high as possible, the vortex state seems less favorable than the single-domain state. Actually, to change the toroidal moment of vortex polarization the curled electric field, $\vec{E}_{cur} = \frac{1}{2}\vec{Q} \times \vec{r}$, originated from a quasi-static magnetic field, $rot\vec{E}_{cur} = -\frac{\partial}{\partial t}\vec{B}$, is required [52, 54]. Corresponding vorticity vector $Q \geq 10^{16}$ V/m$^2$ is very high [78].

However the vortex-like domain structure is relatively sensitive to the screening conditions of ferroelectric polarization; namely it occurs and becomes absolutely stable with $\Lambda/\varepsilon_{IF}$ increase. Thus it makes sense using enough small values of $\Lambda/\varepsilon_{IF}$ to keep the nanoparticle in a stable single-domain state.

Note that Chen and Fang [55] considered ECE in BaTiO$_3$ nanoparticle within core–shell model. Unfortunately the depolarization effects, which are inevitable in the case of zero polarization at the particle surface considered in [55], were completely neglected, and this fact does not allow us applying obtained results to real systems.

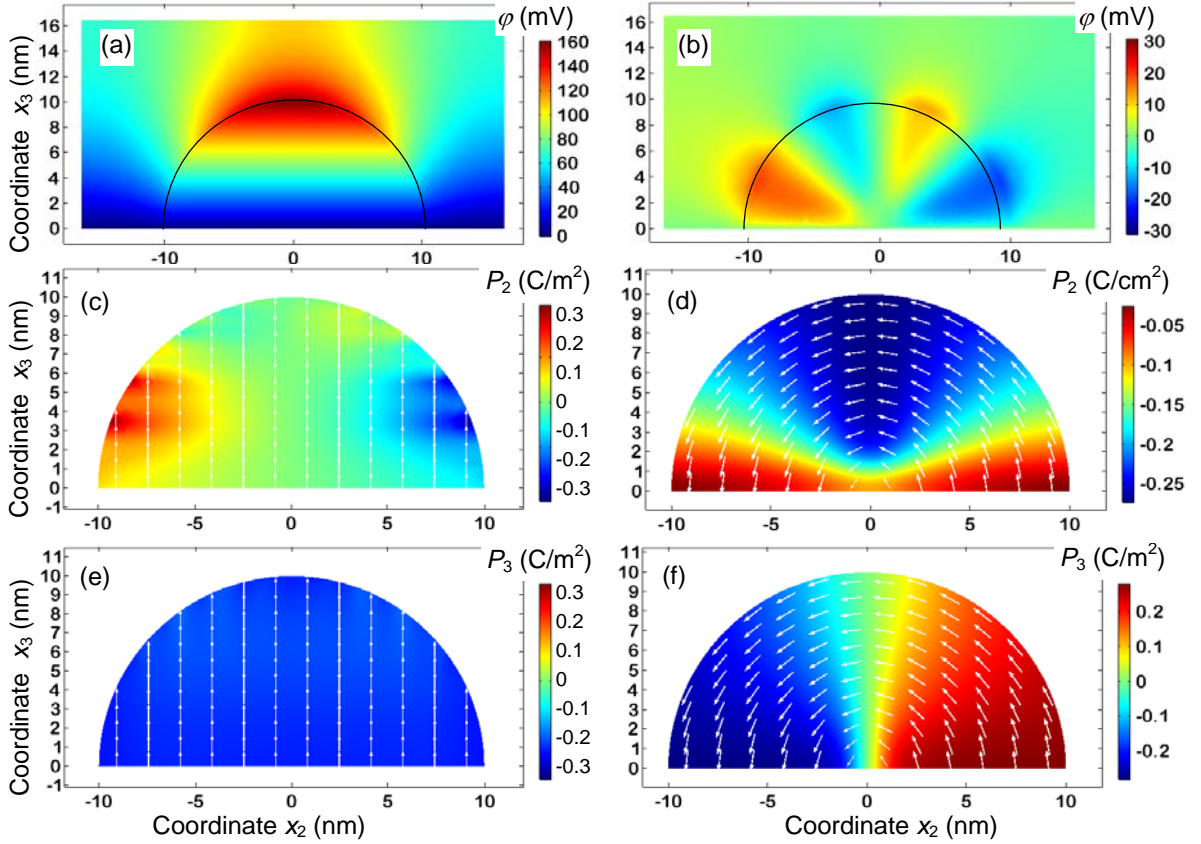

**FIGURE 2**. Distribution of electrostatic potential (**a, b**), polarization components $P_2$ (**c, d**) and $P_3$ (**e, f**) in the cross-section $x_2 = 0$ of the BaTiO$_3$ nanoparticle with radius 10 nm. Two different states are shown, single-domain (**a, c, e**) and vortex-like structure (**b, d, f**). White arrows show polarization direction. Calculations were performed at $T = 293$ K, $\varepsilon_{IF} = 300$, $\Lambda = 2$ nm, $\varepsilon_e = 15$. BaTiO$_3$ parameters are listed in **Table I** for the left plots (**a, c, e**) and in **Table AI** in **Appendix A** for the right plots (**b, d, f**).

### III. APPROXIMATE ANALYTICAL SOLUTION

Phase diagrams of spherical ferroelectric nanoparticles covered by a screening charge layer have several phases, namely paraelectric phase, single-domain ferroelectric phase and poly-domain ferroelectric phase including various domain morphologies [26]. Free energy with renormalized coefficients has the form [see **Appendix A**]:

$$g_R = \alpha_T [T - T_{cr}(R,\Lambda)] \frac{P^2}{2} + \beta(T) \frac{P^4}{4} + \gamma(T) \frac{P^6}{6} - \eta P E_{ext}, \qquad (5a)$$

where $E_{ext}$ is the external electric field and $\overline{P}_3 \equiv P$. Approximate expression for the nanoparticle transition temperature $T_{cr}$ from the single-domain ferroelectric to paraelectric phase is

$$T_{cr}(R,\Lambda) = T_C^* - \frac{1}{\alpha_T \varepsilon_0 [\varepsilon_b + 2\varepsilon_e + \varepsilon_{IF}(R/\Lambda)]}, \qquad (5b)$$

Here the first term $T_C^*$ is the Curie temperature (possibly renormalized by the surface stress). The second term originates from a depolarization field [26].

In Eq.(5a) we introduced the screening factor of the external field,

$$\eta(R,\Lambda) = \frac{3\varepsilon_e}{\varepsilon_b + 2\varepsilon_e + \varepsilon_{IF}(R/\Lambda)}. \quad (5c)$$

Derivation of Eq.(5) is given in **Appendix A.** Note that the expression (5) is exact for the natural boundary conditions for polarization at the particle surface.

Minimization of the free energy Eqs.(5) leads to the equation for the polarization $\alpha_T(T - T_{cr})P + \beta P^3 + \gamma P^5 = \eta E_{ext}$. Differentiation of the equation over external field and temperature leads to the equations $(\alpha_T(T - T_{cr}) + 3\beta P^2 + 5\gamma P^4)\left(\frac{dP}{dE_{ext}}\right)_T = \eta$ and $\left(\frac{\partial P}{\partial T}\right)_E [\alpha_T(T - T_{cr}) + 3\beta P^2 + 5\gamma P^4] = -\alpha_T P - \beta_T P^3 - \gamma_T P^5$, respectively. Using these equations we derived analytical expressions for PE coefficient and EC temperature change (4):

$$\Pi(R,\Lambda) = \frac{\alpha_T P + \beta_T P^3 + \gamma_T P^5}{\alpha_T[T - T_{cr}(R,\Lambda)] + 3\beta P^2 + 5\gamma P^4}, \quad (6a)$$

$$\Delta T_{EC} = \frac{T}{\rho}\int_{E_1}^{E_2}\frac{\Pi(R,\Lambda,P(E))}{C_P}dE = \frac{T}{\eta\rho}\int_{P_1}^{P_2}\frac{\alpha_T P + \beta_T P^3 + \gamma_T P^5}{C_P}dP$$
$$\approx \frac{T}{\eta\rho C_P}\left(\frac{\alpha_T}{2}[P^2(E_2) - P^2(E_1)] + \frac{\beta_T}{4}[P^4(E_2) - P^4(E_1)] + \frac{\gamma_T}{6}[P^6(E_2) - P^6(E_1)]\right) \quad (6b)$$

Approximate equality in the expression (6b) is valid for the case of a ferroelectric with the linearly temperature dependent LGD-expansion coefficients (e.g. for BaTiO$_3$), and for a negligibly weak field dependence of the specific heat, that may be a rough approximation for vast majority of ferroelectric perovskites. More rigorously,

$$C_P \approx \begin{cases} C_P^0 + \dfrac{T(\alpha_T P + \beta_T P^3 + \gamma_T P^5)^2}{\alpha_T(T - T_{cr}) + 3\beta P^2 + 5\gamma P^4}, & T < T_\theta, \\ C_P^0, & T > T_\theta, \end{cases} \quad (7)$$

where $T_\theta = T_{cr} + \dfrac{\beta^2}{4\gamma\alpha_T}$ is the maximal temperature of the ferroelectric phase metastability. Derivation of Eq.(7) utilizes Eq.(4b) and the fact that the g-derivatives simplifies allowing for the equation $\alpha_T(T - T_{cr})P + \beta P^3 + \gamma P^5 = \eta E_{ext}$ (see **Appendix B**). Following Landau theory, the dielectric susceptibility $\chi_E = \dfrac{1}{\alpha_T(T - T_{cr}) + 3\beta P^2 + 5\gamma P^4}$ diverges at $T = T_\theta$ and $E_{ext} = 0$, while the polarization is finite for the 1$^{st}$ order phase transitions, leading to the divergence of the difference $\delta C_P = C_P - C_P^0$. In reality both the external electric field and critical fluctuations transform the divergence into a maximum that is typically $\approx 10 - 30$ % in height of $C_P$ (however there can be exceptions). Typically the maximum shape cannot be described by a rigorous analytical expression, but semi-empirically as $\chi_E = \dfrac{1}{\sqrt{(\alpha_T(T-T_{cr})+3\beta P^2+5\gamma P^4)^2+\delta^2}}$, where the empirical parameter $\delta$ is small enough.

Since the polarization-dependent term $\dfrac{T(\alpha_T P+\beta_T P^3+\gamma_T P^5)^2}{\alpha_T(T-T_{cr})+3\beta P^2+5\gamma P^4}$ is positive, it always increases $C_P$ and so decreases the integrand expression in Eq.(6b). As a result, the approximate expression (6b) overestimates the ECE.

If the dimensionless parameter $\mu = \dfrac{T\alpha_T^3}{C_P^0 \beta_T}\overline{\chi}_E$ is small, the first-order corrections to Eq.(6b) have the form:

$$\Delta T_{EC} \approx \dfrac{T}{\eta\rho C_P^0}\left(\begin{array}{l}\dfrac{\alpha_T}{2}\left[P^2(E_2)-P^2(E_1)\right]+\dfrac{\beta_T(1-\mu)}{4}\left[P^4(E_2)-P^4(E_1)\right]\\ +\dfrac{\gamma_T-3\mu\alpha_T^{-1}\beta_T^2}{6}\left[P^6(E_2)-P^6(E_1)\right]+....\end{array}\right) \qquad (8)$$

In the linear approximation, valid for weak enough external fields (i.e. for $E_{ext}$ much lower than the coercive field), $P(E_{ext}) \approx P(0) - \dfrac{\eta E_{ext}}{2\alpha_T(T-T_{cr})}$, and so $P^2(E_{ext}) - P^2(0) \approx -\dfrac{\eta E_{ext} P_S}{\alpha_T(T-T_{cr})}$, where $P(0) \equiv \pm P_S$ is the nanoparticle spontaneous polarization. Within the approximation

$$\Delta T_{EC}(E_{ext}) \approx -\dfrac{T}{\rho C_P^0}\left(\alpha_T P_S+\beta_T P_S^3+\gamma_T P_S^5\right)\dfrac{E_{ext}}{2\alpha_T(T-T_{cr})}, \qquad (9)$$

and the nanoparticle spontaneous polarization is:

$$P_S(R,\Lambda,T) = \sqrt{\dfrac{1}{2\gamma}\left(\sqrt{\beta^2+4\gamma\alpha_T(T_{cr}(R,\Lambda)-T)}-\beta\right)} \qquad (10)$$

## IV. SIZE EFFECT ON PYROELECTRIC AND ELECTROCALORIC PROPERTIES

Below we analyze the correlations between the nanoparticle polarization $P$, relative dielectric permittivity $\varepsilon_{NP}$, PE coefficient $\Pi$, and EC temperature change $\Delta T_{EC}(E)$ calculated for a periodic external electric field, $E_{ext}=E_0\sin(\omega t)$, different temperature, $T$, and nanoparticle radius, $R$. **Figures 3-6** show typical dependences of $P$, $\varepsilon_{NP}$, $\Pi$ and $\Delta T_{EC}$ on $E_{ext}$, $T$, and $R$ for BaTiO$_3$ nanoparticles with parameters listed in **Table I**. All dependences are calculated for the relatively high interfacial permittivity $\varepsilon_{IF}=300$ (that is realistic for paraelectric shells), enough high effective screening length $\Lambda=2$ nm, and ambient permittivity $\varepsilon_e=15$ characteristic for a high-k dielectric matrix (e.g. widely used PVDF). We compared the "static" dependences (dashed curves), which include unstable and metastable regions, with thermodynamically stable "dynamic" dependences (solid curves) calculated for the external field frequency $\omega = 2 \times 10^4$ s$^{-1}$.

Dependences of $P$, $\varepsilon_{NP}$, $\Pi$, and $\Delta T_{EC}$ on $E_{ext}$ are shown in **Figs. 3a-d**, respectively. The dependences are calculated for several nanoparticle radii $R$ (curves 1 - 4) at room temperature. The ferroelectric hysteresis loop $P(E_{ext})$ is narrow for the smallest particle ($R = 4$ nm), then it expands and becomes significantly wider (i.e. the coercive field $E_C$ increases) with the increasing particle radius (compare solid curves 1 - 4 in **Fig. 3a**). Note, that the appearance of a very narrow hysteresis loop at $R = 4$ nm is a purely dynamic effect observed at nonzero frequency $\omega$. Actually, the static dashed black curve calculated for $R = 4$ nm does not contain any unstable S-shaped region. Other static curves calculated for $R > 5$ nm contain the unstable S-shaped region corresponding to the bistable states of the ferroelectric polarization. All curves and loops in **Fig. 3a** show the behavior typical for the ferroelectric nanoparticles undergoing the 1$^{st}$ order FEPT to a paraelectric phase with $R$ decrease (i.e. size-induced phase transition). From Eq. (5b) the critical radius $R_{cr}$ of the size-induced transition is given by expression,

$$R_{cr}(T, \Lambda) = \Lambda \left( \frac{1}{\alpha_T \varepsilon_0 \varepsilon_{IF}(T_C^* - T)} - \frac{\varepsilon_b + 2\varepsilon_e}{\varepsilon_{IF}} \right). \qquad (11)$$

From Eq.(11) the critical size is about 8 nm at 293 K and $E_0 = 0$, and so the particle with $R = 4$ nm is paraelectric, and other with $R = (10 – 20)$ nm are ferroelectric at room temperature.

Correlating with **Fig. 3a**, **Figs. 3b-c** illustrate the characteristic features (maximum, sharp double maximum, or divergence) of $\varepsilon_{NP}$ and $\Pi$ emerging in the vicinity of the coercive field that value increases with the $R$ increase. Maxima correspond to nonzero frequency $\omega > 0$, and divergences are for $\omega = 0$. The "unphysical" values of negative permittivity corresponding to the unstable S-shaped regions in **Fig. 3a** (dashed curves) are not shown in **Fig. 3b**.

The dependences $\Delta T_{EC}(E_{ext})$ calculated at $\omega \neq 0$ (solid curves in **Fig. 3d**) correlate with the dependences $\varepsilon_{NP}(E_{ext})$ (solid curves in **Fig. 3b**), but have several distinctive features. For the smallest "paraelectric" particles ($R \leq 4$ nm) the $\Delta T_{EC}(E_{ext})$ static dependences (dashed curves) and very narrow dynamic loops (solid curves) have a vase shape (without maximums). The $\Delta T_{EC}$ value monotonically increases and then saturates with $E_{ext}$ increasing (solid curve 1 in **Fig. 3d**). In the "paraelectric" phase ECE is positive ("heating" effect). It should be mentioned, that such behavior correlates with quadratic field dependence of the electrocaloric effect in bulk paraelectric [79].

For the bigger "ferroelectric" particles ($R \geq 10$ nm) the $\Delta T_{EC}(E_{ext})$ dependences are hysteretic (solid curves 2-4 in **Fig. 3d**). The ECE changes its sign to negative ("cooling" effect) when the electric field becomes antiparallel to the polarization direction. Near the coercive field $\Delta T_{EC}$ reaches a pronounced maximum and changes the sign back to the positive one just above it. The ECE maxima sharpness and their magnitude increases with the increasing particle radius. In the ferroelectric phase (for $R \geq 10$ nm) ECE is relatively small with the exception of the coercive field vicinity, where it reaches minus 2-3 K. The static $\Delta T_{EC}(E_{ext})$ dependences (**Fig. 3d**, dashed curves 2-4) almost coincide with dynamic those (**Fig. 3d**, solid curves 2-4), except for the instability region marked by a dotted rectangle (see **Fig. 3d**, bottom).

As can be seen from the **Fig. 3c,d**, one can induce the appearance, control the width, magnitude and sign of $\Pi(E)$ and $\Delta T_{EC}(E)$ maxima by changing the particle size, as well as tune the field interval, within which PEE and ECE are maximal.

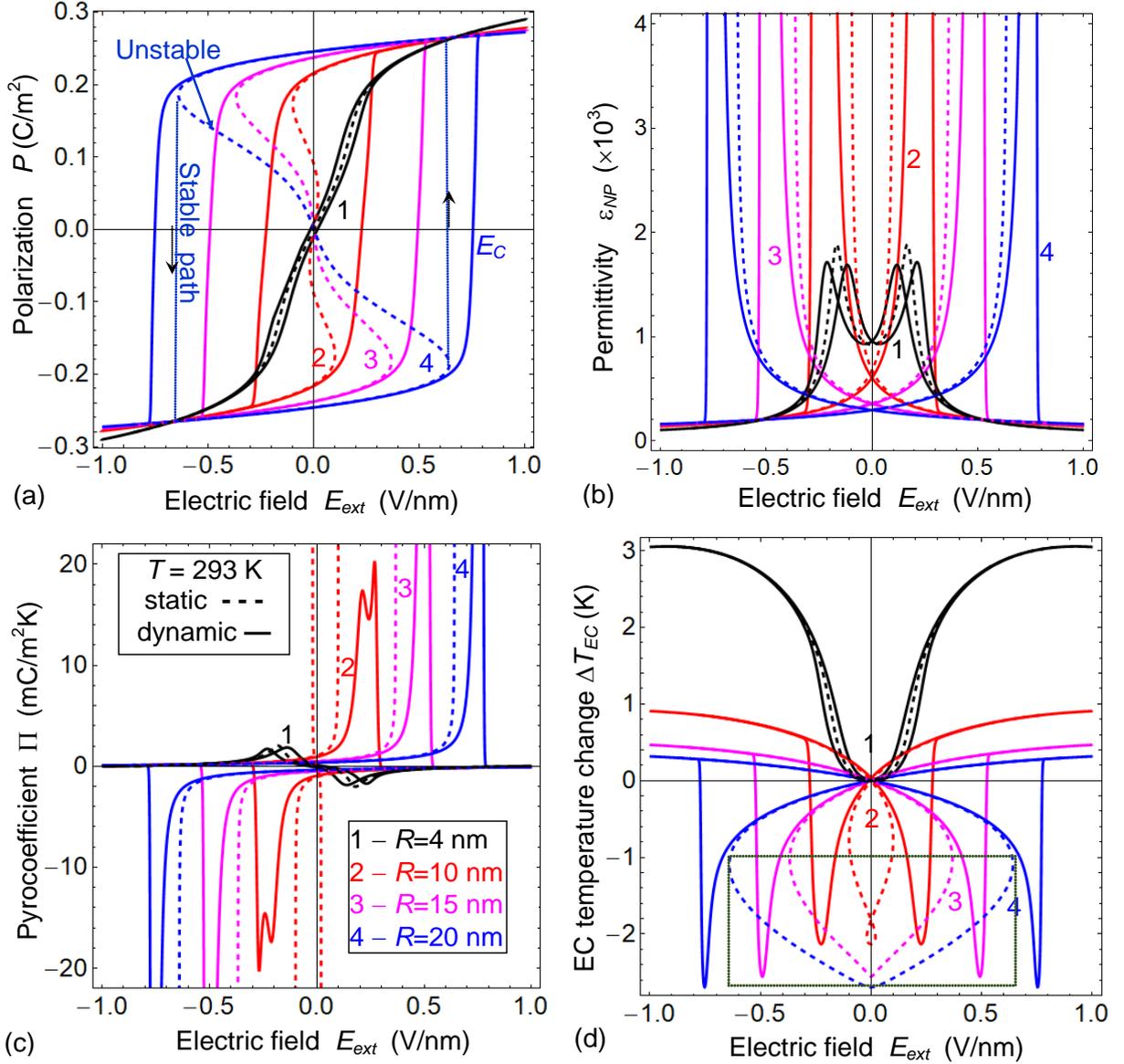

**FIGURE 3.** Dependences of the polarization (**a**), relative dielectric permittivity (**b**), PE coefficient (**c**) and EC temperature change (**d**) on external electric field calculated for several radii $R = 4, 10, 15, 20$ nm of BaTiO$_3$ nanoparticle (curves 1-4), $T = 293$ K, $\varepsilon_{IF} = 300$, $\Lambda = 2$ nm, $\varepsilon_e = 15$, $\Gamma = 10^2$ SI units, and $\omega = 2 \times 10^4$ s$^{-1}$ for solid curves. The static dependences ($\omega = 0$) including unstable regions are shown by dashed curves. Dotted vertical lines with arrows in plot (a) show stable paths. The unstable dependences are shown by dashed curves inside the dotted rectangle in plot (d). BaTiO$_3$ parameters are listed in **Table I**, its density $\rho = 6.02 \times 10^3$ kg/m$^3$ and specific heat $C_p = 4.6 \times 10^2$ J/(kg·K) at room temperature.

Temperature dependences $P(T)$, $\varepsilon_{NP}(T)$, $\Pi(T)$ and $\Delta T_{EC}(T)$, calculated for the several radii of nanoparticles and a relatively small amplitude of external field ($E_0 = 0.01$ V/nm) in the vicinity of the FEPT region, are shown in **Fig. 4a-d**. The amplitude $E_0$ is well below the thermodynamic coercive field of polarization reversal in our case (0.2 – 0.8 V/nm) **Fig.3a**. The temperature "hysteresis", defined as the interval between the solid and dotted vertical lines, is the widest for the smallest $R$ and narrows when the particle radius increases (compare solid and dashed curves 1-4 in **Fig. 4a**). The origin of the hysteresis is thermodynamic bistability, therefore it disappears at bigger $E_0$ and/or for higher ω. The hysteresis position corresponds to the vicinity of the transition temperature, $T_{cr}$, which is particle size dependent and goes up with increasing $R$, in agreement with Eq. (5b).

In fact, all static dependences in **Fig. 4** (dashed curves 1 – 4) contain the unstable S-shaped region, which width decreases with the increasing particle size. The dependences $P(T)$ in **Fig. 4a** show the behavior typical for the ferroelectric nanoparticle undergoing the 1$^{st}$ order FEPT below $T_{cr}$. Correlating with **Fig. 4a**, **Figs. 4b-c** show typical sharp maxima at ω > 0 on $\varepsilon_{NP}(T)$ and $\Pi(T)$ or their divergence for ω = 0 emerging at $T_{cr}(R)$. The "unphysical" regions of the negative permittivity, corresponding to the unstable "inverse S"-shaped regions at the dashed curves in **Fig. 4a** are not shown in **Fig. 4b**.

Temperature dependences of EC temperature change, $\Delta T_{EC}(T)$, calculated at frequency ω ≠ 0 (shown by solid curves in **Fig. 4d**) reveal several distinct features. $\Delta T_{EC}(T)$ sharply increases from almost zero values and reaches several Kelvins in the region of the $P(T)$ hysteresis. Correlating with the $P(T)$ behavior at FEPT (shown in **Fig. 4a**), maximal $\Delta T_{EC}$ is located around $T_{cr}$. The magnitude of the $\Delta T_{EC}$ maximum increases and its width decreases with the increasing particle size (see solid curves 1-4 in **Fig. 4d**). The region of nonzero ECE becomes broader with the decreasing ω (compare the distance between solid, dashed, and dotted vertical lines). Note that the static dependences contain unstable regions, which width decreases with increasing $R$ (see dashed curves 1-4 in **Fig. 4d**). As can be seen from the **Fig. 4c,d**, one can control the width and height of $\Pi(T)$ and $\Delta T_{EC}(T)$ peaks by changing the particle size, as well as select the temperature interval where $\Pi(T)$ and $\Delta T_{EC}(T)$ are maximal. This conclusion is valid for small $E_{ext}$ amplitude.

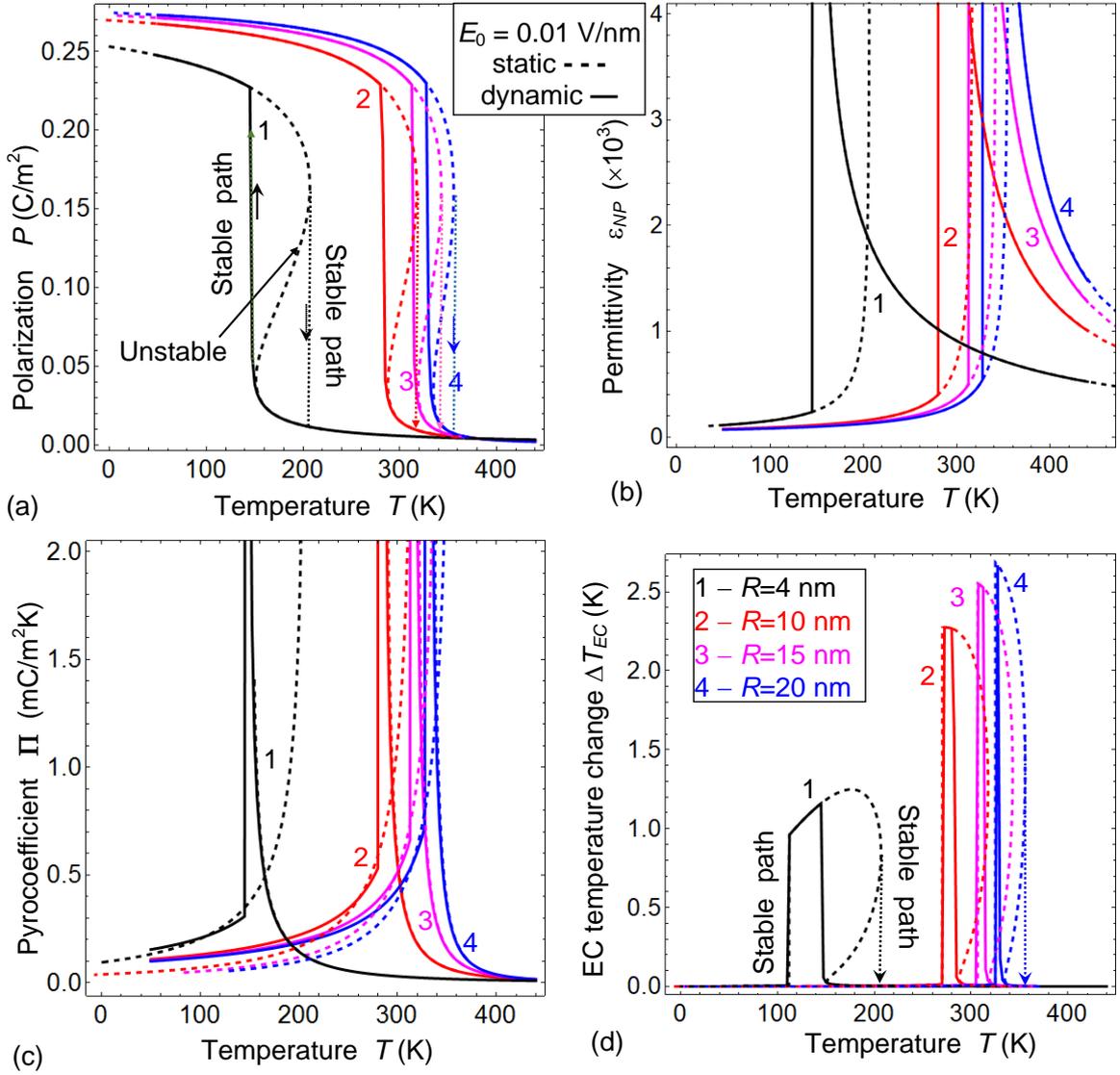

**FIGURE 4.** Temperature dependences of polarization (a), relative dielectric permittivity (b), PE coefficient (c) and EC temperature change (d) calculated for several radii of BaTiO$_3$ nanoparticle $R$ = 4, 10, 15, 20 nm (curves 1-4) and external electric field amplitude $E_0$ = 0.01 V/nm. The field frequency $\omega = 2\times10^4$ s$^{-1}$ for solid curves, and $\omega = 0$ for dashed curves, which include unstable regions. Dotted vertical lines with arrows show stable paths. Other parameters are the same as in **Fig. 3**.

The dependences $P(T)$, $\varepsilon_{NP}(T)$, $\Pi(T)$ and $\Delta T_{EC}(T)$ calculated for several nanoparticle radii at rather large $E_{ext}$ amplitude of ($E_0$ = 0.5 V/nm) and relatively low frequency $\omega = 2\times10^4$ s$^{-1}$ are shown in **Figs. C1a-d, Appendix C**. The temperature hysteresis, existing for small $E_0$, disappears and all dependences are significantly "smeared" with $E_0$ increasing. Rather asymmetric maxima of the $\varepsilon_{NP}(T)$ and $\Pi(T)$ emerge at the phase transition temperature, but their temperature position is almost radius-independent. At large $E_0$ the dependences $\Delta T_{EC}(T)$ reveal several distinctive features (compare **Fig. C1d** with **Fig. 4d**). First, for all particle sizes the maximum of $\Delta T_{EC}(T)$ lies definitely below the peak of the dielectric permittivity. Second, the temperature range of the maximal $\Delta T_{EC}$ narrows and
15

shifts to higher temperatures when the particle radius increases due to FEPT sharpening and the shift of $T_{cr}$. Note that even at the strong fields we still can control the temperature range and width of the maximal PEE and ECE by changing the particle size.

The static dependences $P(T)$, $\varepsilon_{NP}(T)$, $\Pi(T)$ and $\Delta T_{EC}(T)$ calculated for several values of the constant external field $E_{ext}=E_0$ and $R = 10$ nm are shown in **Figs. 5a-d**. The temperature hysteresis of $P(T)$ existing for small $E_0$ (shown by black dotted lines in **Fig. 5a**) disappears, and the dependence $P(T)$ becomes significantly smeared with $E_0$ increasing above a critical field $E_{cr}$ that is about 0.1 V/nm for $R = 10$ nm. (compare solid curves 1-4 in **Fig. 5a**). Correlating with **Fig. 5a**, **Figs. 5b-c** show rather asymmetric maxima of the $\varepsilon_{NP}(T)$ and $\Pi(T)$ emerging at the phase transition temperature, which magnitude and sharpness noticeably decreases, and the position shifts to the higher temperatures with increasing $E_0$. Such behavior is typical for smearing of the 1$^{st}$ order FEPT region by an external field.

The dependences $\Delta T_{EC}(T)$ for different $E_0$ are shown in **Fig. 5d**. For larger $E_0$ the $\Delta T_{EC}$ value is nonzero, but rather small, at $T < 260$ K, and sharply increases at 260 K. This temperature of a sudden rise of the ECE does not depend on the electric field value. At higher temperatures $\Delta T_{EC}$ is rising, reaches a maximum, which positions shifts towards higher temperatures on increasing of the electric field, and then gradually decreases (see curves 2-4 in **Fig. 5d**). For $E_0 = 5 \cdot 10^8$ V/m the maximal $\Delta T_{EC} \approx 3$ K occurs in vicinity of 360 K. The temperature range of nonzero $\Delta T_{EC}$ significantly broadens with increasing $E_0$ (compare the shape of maxima for the curves 1-4 in **Fig. 5d**), which correlates with the $P(T)$ behavior (curves 1-4 in **Fig. 5a**) due to FEPT smearing at high electric field. Note that the temperature hysteresis of $\Delta T_{EC}(T)$ existing for small $E_0$ [shown by black curves in **Fig.5d**] disappears with increasing $E_0$.

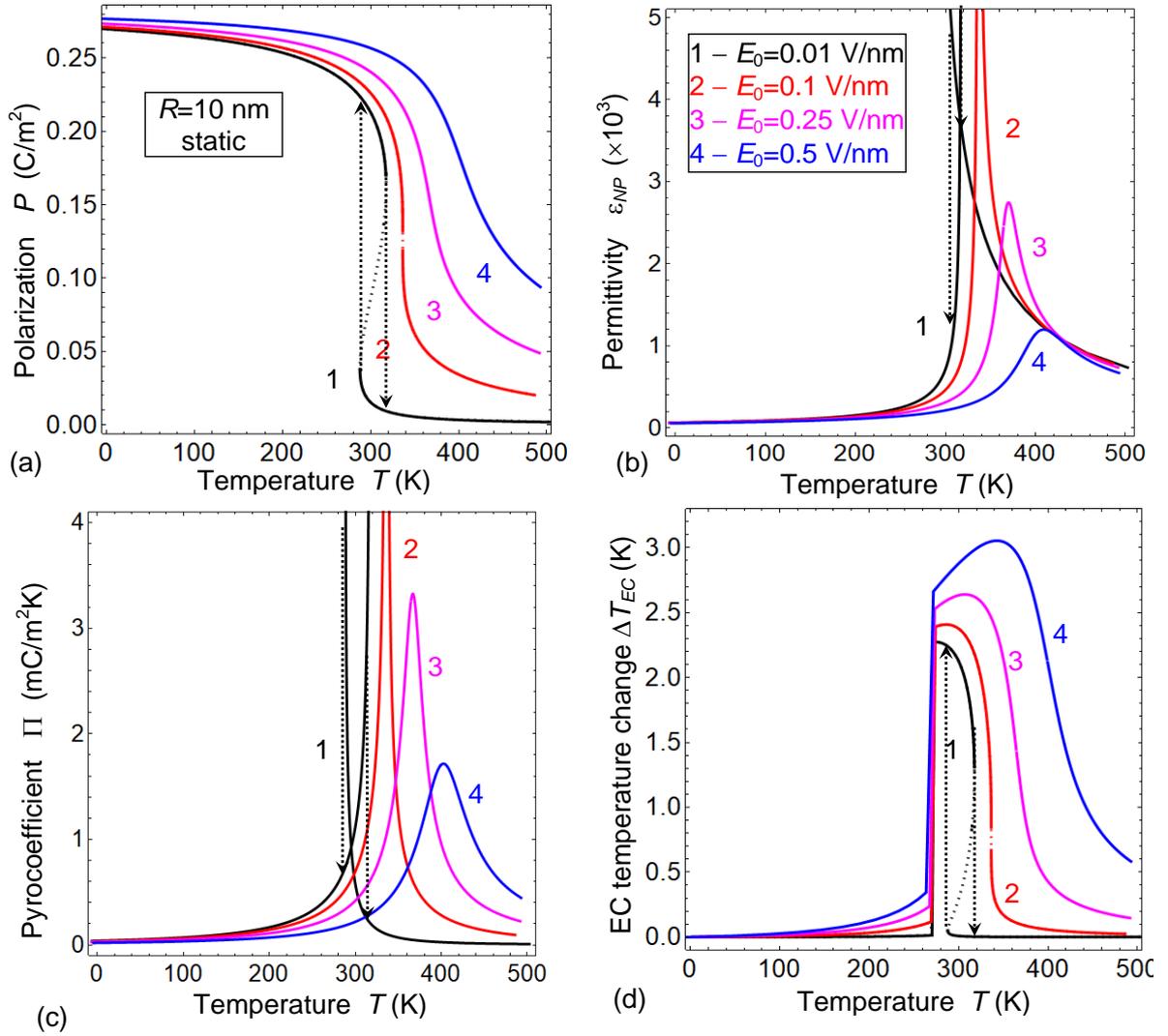

**FIGURE 5.** Static temperature dependences of the BaTiO$_3$ nanoparticle polarization **(a)**, relative dielectric permittivity **(b)**, PE coefficient **(c)** and EC temperature change **(d)** calculated for BaTiO$_3$ nanoparticle with radius of $R = 10$ nm and different external field $E_0 = 0.01, 0.1, 0.25, 0.5$ V/nm (curves 1-4). Other parameters are the same as in **Fig. 3**.

Particle radius dependences $P(R)$, $\varepsilon_{NP}(R)$, $\Pi(R)$, and $\Delta T_{EC}(R)$ calculated for several amplitudes of the external field, $E_0$ and $T = 293$ K are shown in **Fig. 6a-d**. These dependences correlate with the dependences $P(T)$, $\varepsilon_{NP}(T)$, $\Pi(T)$ and $\Delta T_{EC}(T)$ shown in **Fig. 4a-d**, since $T_{cr} \sim 1/R$ per Eqs.(5b) and (8). The "size hysteresis", defined as the distance (in nm) between solid and dotted vertical lines, is the widest for the smallest $E_0$, and it narrows and disappears when $E_0$ increases above the critical value $E_{cr}$, that is about 0.1 V/nm at 293 K (compare solid and dashed curves 1-4 in **Fig. 6a**). The origin of the effect is a thermodynamic bistability, and so it disappears with the frequency increase above the critical value, that is temperature- and field- dependent. The bistability region corresponds to the vicinity of the temperature-dependent critical radius $R_{cr}$ given by Eq. (11). The static dependences for small $E_0$ contain the unstable S-shaped regions, which width strongly decreases with $E_0$ increase (compare solid and dashed curves 1-2 in **Fig. 6a**). The dependences shown in **Fig. 6a** illustrate the scenario of the 1$^{st}$ order size-induced FEPT. Correlating with **Fig. 6a**, **Fig. 6b-c** show sharp maxima of $\varepsilon_{NP}(R)$ and $\Pi(R)$ emerging at the critical radius. With increasing $E_0$ the width of the $\varepsilon_{NP}(R)$ peak significantly decreases (curves 1-3 in **Fig. 6b**), while the width of the $\Pi(R)$ peak remains almost unchanged (curves 1-3 in **Fig. 6c**). These peaks disappear for the strong enough field $E_0$ (curves 4 in **Fig. 6d,c**).

Particle radius dependences $\Delta T_{EC}(R)$ calculated at $T = 293$ K, low frequency ($\omega = 2\times10^4$ s$^{-1}$), and several amplitudes $E_0$ are shown by solid curves in **Fig. 6d**. They look very different for small ($E_0 < 0.1$ V/nm) and big ($E_0 \gg 0.1$ V/nm) amplitudes of $E_{ext}$ (compare curves 1-3 with curve 4 in **Fig. 6d**). The $\Delta T_{EC}(R)$ behavior strongly correlates with the $P(R)$ behavior shown in **Fig. 6a**. In particular, the size hysteresis of $\Delta T_{EC}(R)$ disappears for $E_0$ larger than the critical value $E_{cr} \approx 0.1$ V/nm. For $E_0 \leq 0.1$ V/nm $\Delta T_{EC}$ first increases with $R$ from negligibly low values to $\approx 2.5$ K in the region of polarization hysteresis and then abruptly drops back to very small values above the critical radius (see curves 1-3 in **Fig. 6d**). For $E_0 = 0.5$ V/nm $\Delta T_{EC}$ is large already at the smallest radius, slightly increases from 2.5 K to $\approx 3$ K and then abruptly drops above the critical radius (see curve 4 in **Fig. 6d**) in correlation with $P(R)$ curve 4 in **Fig. 6a**. The range, where the large ECE is observed, depends on frequency and strongly increases with decreasing $\omega$ (compare the interval between solid "dynamic" and dashed "static" curves in **Fig. 6d**). Note that the static curves contain the unstable regions, which width strongly decreases with the $E_0$ increase (see dashed curves in **Fig. 6d**). The region of nonzero static $\Delta T_{EC}(R)$ is the widest for the largest $E_0$ and becomes significantly narrower with $E_0$ decrease. As can be seen from the **Fig. 6d**, we can control the shape, magnitude and width of $\Delta T_{EC}(R)$ peak by changing the electric field amplitude.

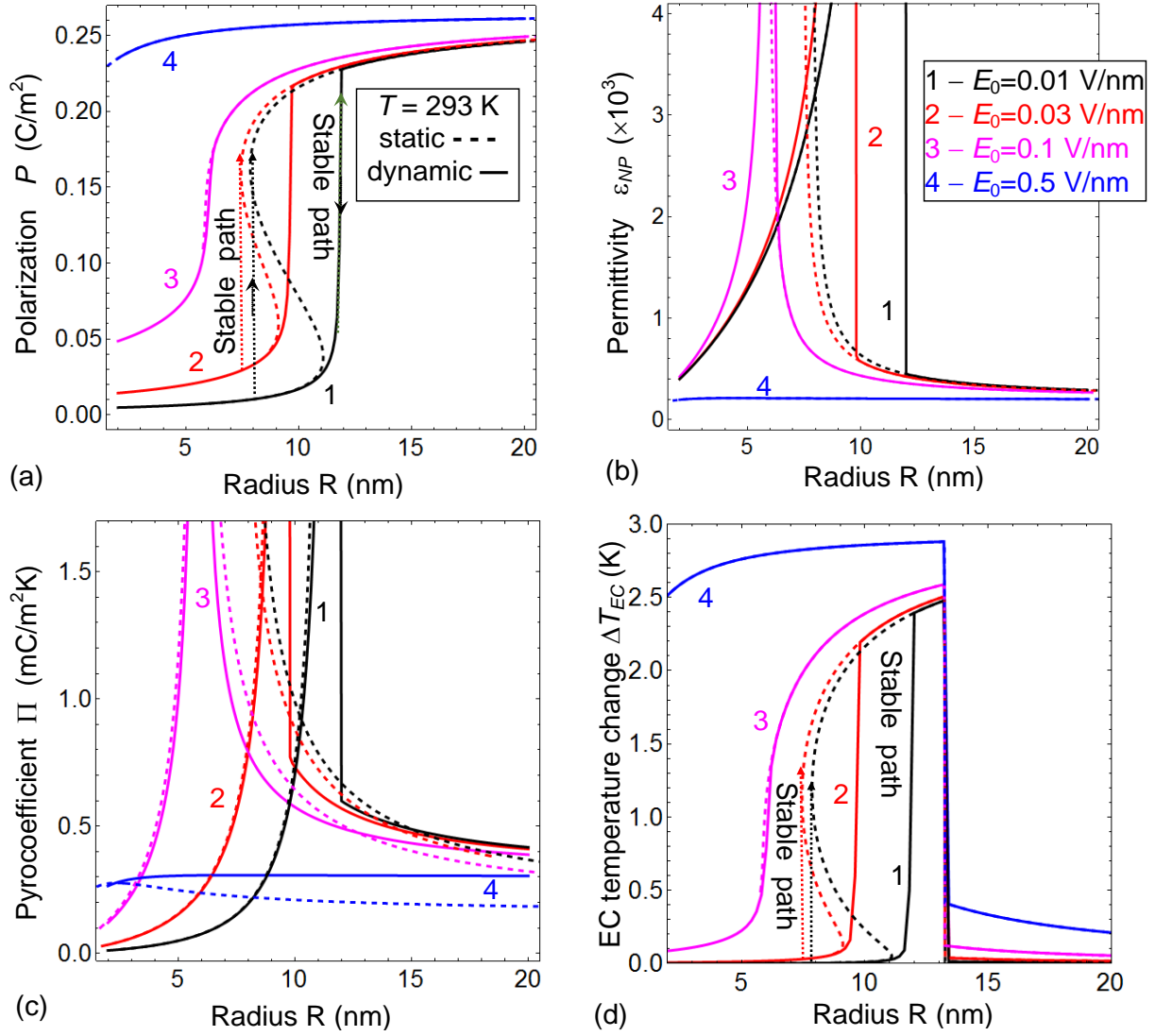

**FIGURE 6.** Radius dependences of the BaTiO$_3$ nanoparticle polarization **(a)**, relative dielectric permittivity **(b)**, PE coefficient **(c)** and EC temperature change **(d)** calculated at T = 293 K for several amplitudes of external field $E_0$ = 0.01, 0.03, 0.1, 0.5 V/nm (curves 1-4), frequency $\omega = 2 \times 10^4$ s$^{-1}$ (solid curves) and $\omega = 0$ (dashed curves). Other parameters are the same as in **Fig. 3**.

## V. SIZE EFFECT OF PYROELECTRIC AND ELECTROCALORIC ENERGY CONVERSION

### A. Size effect of pyroelectric figures of merit

For better displaying pyroelectric energy conversion it is convenient to consider the corresponding figures of merit (**FoM**) of pyroelectric materials. According to the operation modes of pyroelectric convertors [41, 80, 81, 82] the current ($F_I$) and voltage ($F_V$) FoM have been introduced:

$$F_I = \frac{\Pi}{c_P}, \quad F_V = \frac{\Pi}{\varepsilon_0 \varepsilon c_P}. \tag{12a}$$

Here $c_P = \rho C_P$ is the volume heat capacity and $\rho$ is the density of the PE material.

In the energy conversion operation mode, a PE convertor of the capacity type generates a pyro-charge $Q_\pi$ during a thermal cycle. At that, the electric energy generated during the heating/cooling cycle is proportional to $Q_\pi^2$. If the PE convertor is subjected to an incident radiation, the electric energy generated during the thermal cycle is proportional to the square of the pyro-voltage $U_\pi^2$. For both these cases two different energy conversion FoM, $F_{EQ}$ and $F_{EU}$, have been proposed [83]:

$$F_{EQ} = \frac{\Pi^2}{\varepsilon_0 \varepsilon}, \quad F_{EU} = \frac{\Pi^2}{\varepsilon_0 \varepsilon c_P^2} \quad (12b)$$

The efficiency of the PE energy conversion is defined by the pyroelectric (electro-thermal) coupling factor [82, 83]:

$$k_{PE}^2 = \frac{\Pi^2 T}{c_P \varepsilon_0 \varepsilon}, \quad (12c),$$

where $T$ is the ambient temperature. Some details of derivation of expressions (9b-c) is given in **Appendix D.**

Expressions (9) for the PE FoM and coupling factor are valid for a freely suspended ferroelectric layer, and should be modified for nanocomposites, hybrid, or/and layered nanosystems. In accordance with the theory of finite size effects in ferroelectric nanomaterials [84], the form of basic expressions relatively often remains unchanged, but the parameters are substituted by effective ones. Hence, we introduce the PE FoMs and coupling constant for nanoparticles (**NP**) in the following form:

$$F_I = \frac{-\Pi}{c_{NP}}, \quad F_f = \frac{-\Pi}{\varepsilon_0 \varepsilon_{NP}}, \quad K_{PE} = \frac{\Pi^2}{\varepsilon_0 \varepsilon_{NP} c_{NP}}, \quad F_{EQ} = \frac{\Pi^2}{\varepsilon_0 \varepsilon_{NP}}, \quad F_{EU} = \frac{\Pi^2}{\varepsilon_0 \varepsilon_{NP} c_{NP}^2}. \quad (13)$$

The nanoparticle volume heat capacity, $c_{NP} = \rho C_P^{NP}$, introduced in Eqs.(13), is a temperature- and size- dependent quantity. **Figures 7-8** present temperature dependences of the values (13) calculated for different particle radii $R$ and external electric field amplitude $E_0$.

Static temperature dependences of the specific heat variation $\delta C_P \equiv C_P - C_P^0$, PE FoMs $F_I$, $F_f$, $F_{EQ}$, $F_{EU}$ and coupling constant $K_{PE}$ calculated for several $R$ and relatively small $E_0 \ll E_{cr}$ are shown in **Figs. 7a-f**. Dotted vertical lines with arrows show thermodynamically stable paths. The hysteresis region of the temperature dependences, defined as the distance between the two dashed lines, decreases and shifts towards higher temperatures with increasing the particle radius. It should be noted that the positions of the hysteresis for $\delta C_P$ and all FoM are the same and completely coincide with the position of the $P(T)$ hysteresis in **Fig. 4a.** $F_f(T)$ shows a maximum within the temperature hysteresis region, which magnitude does not depend on the particle size (**Fig. 7c**). This is related to similar character of $\Pi(T)$ and $\varepsilon_{NP}(T)$ divergences (**Fig. 4b,c**). Temperature dependences of other parameters (**Figs. 7b-f**) have either divergences or very sharp maxima at the edges of the temperature hysteresis region, which are slightly suppressed for $F_I$, $K_{PE}$ and $F_{EU}$ due to increased $\delta C_P$ in the region (**Fig. 7a**).

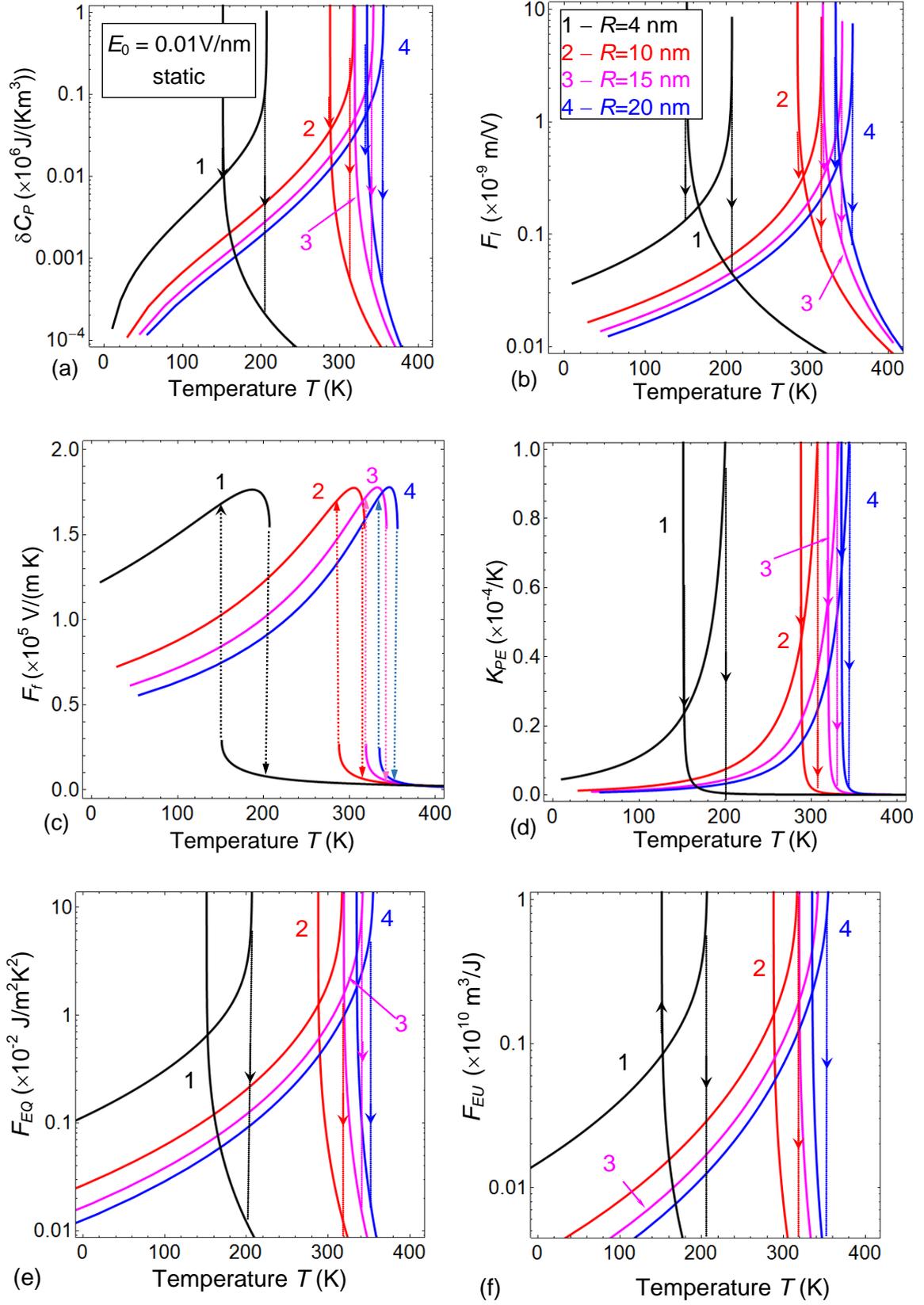

**FIGURE 7.** Static temperature dependences of specific heat variation $\delta C_P \equiv C_P - C_P^0$ (**a**), and PE performances $F_I$ (**b**), $F_f$ (**c**), $K_{PE}$ (**d**), $F_{EQ}$ (**e**), and $F_{EU}$ (**f**) calculated for several radii of BaTiO$_3$ nanoparticles

$R = 4, 10, 15, 20$ nm (curves 1-4). External field amplitude $E_0 = 0.01$ V/nm. Dotted vertical lines with arrows show thermodynamically stable paths under the temperature increase or decrease. Other parameters are the same as in **Fig. 3.**

Static temperature dependences of $\delta C_P$, $F_I$, $F_f$, $K_{PE}$, $F_{EQ}$ and $F_{EU}$ calculated for several amplitudes of the external field $E_0$ and $R = 10$ nm are shown in **Figs. 8a-f**. Dotted curves show thermodynamically unstable regions, where the temperature hysteresis exists. The temperature hysteresis disappear for electric fields above the critical value, $E_0 > E_{cr}$, where $E_{cr}$ is about 0.1 V/nm for $R = 10$ nm.

The temperature dependences of $F_I$, $K_{PE}$, $F_{EQ}$ and $F_{EU}$ (**Figs. 8b,d-f**) have either divergences or very sharp maxima at the edges of temperature hysteresis region. The temperature hysteresis for $F_I$, $K_{PE}$, $F_{EQ}$ and $F_{EU}$ exists at weak fields $E_0 \ll E_{cr}$ (see black curves). The hysteresis region disappears for $E_0 > E_{cr}$ (see curves 2) and corresponding dependences are characterized by the maxima shifted towards higher temperatures under electric field increase (see curves 3-4). The increasing electric field also leads to the decrease of $F_I$, $F_{EQ}$, $F_{EU}$ and $K_{PE}$ maximal values.

The temperature dependence of $F_f$ does not have a divergence even at small $E_0 \ll E_{cr}$, but has a maximum within the temperature hysteresis. It is characterized by the absence of hysteresis for $E_0 > E_{cr}$, and the increase of the electric field leads to the shift of $F_f$ maximum without affecting its value (see curves 2-4 in **Figs. 8c**).

We would like to underline the evident similarity between the temperature dependences of $F_{EQ}$ and $F_{EU}$ (**Figs. 7e,f** and **8e,f**). This similarity originates from the proportionality of both $F_{EQ}$ and $F_{EU}$ to $\frac{\Pi^2}{\varepsilon_0 \varepsilon_{NP}}$ [see Eqs.(13)], as well as from the relatively weak temperature dependence of the full heat capacity, $C_P = \delta C_P(T,R) + C_P^0$, since $\delta C_P(T,R) \ll C_P^0$ outside the immediate vicinity of the size-induced transition to the ferroelectric phase. At that external electric field strongly broadens and suppresses the maximum of $\delta C_P(T,R)$ with retention of sharp character of $\delta C_P$ dependence in intermediate temperature range (compare curves 1-4 in **Figs.7a** and **8a**).

To resume the subsection, the results shown in **Figs 7-8** illustrate the possibility to control PE performances using the dependences of corresponding FoM and coupling factor on operating temperature, external electric field, and nanoparticle size.

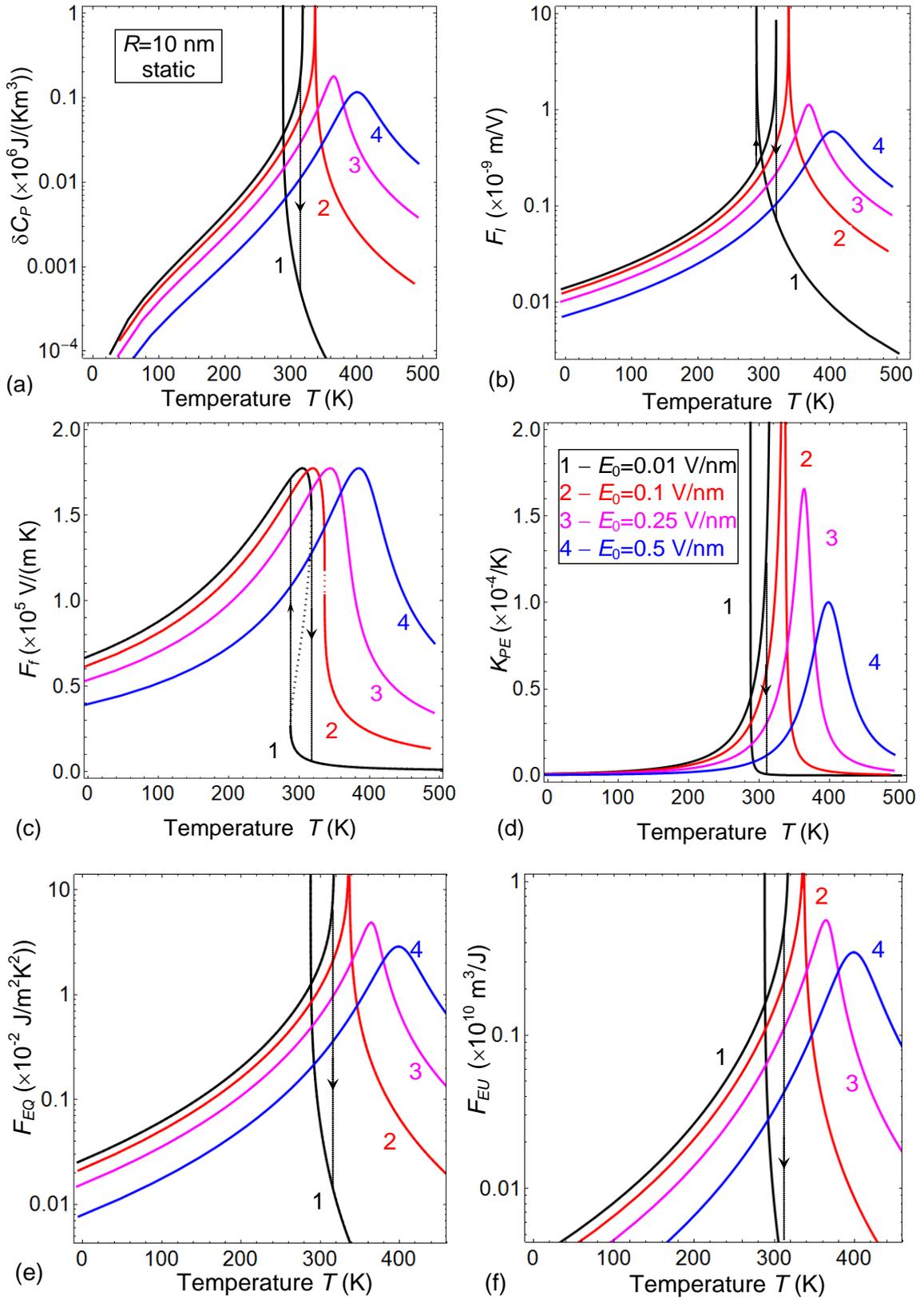

**FIGURE 8.** Static temperature dependences of specific heat variation $\delta C_P \equiv C_P - C_P^0$ **(a)**, and PE performances $F_I$ **(b)**, $F_f$ **(c)**, $K_{PE}$ **(d)**, $F_{EQ}$ **(e)**, and $F_{EU}$ **(f)** calculated for of BaTiO$_3$ nanoparticles with radius $R = 10$ nm and different values of external field $E_0 = 0.01, 0.1, 0.25, 0.5$ V/nm (curves 1-4). Other parameters are the same as in **Fig. 3**.

## B. Size effect and frequency of electrocaloric coefficient hysteresis

Let us study the features of EC coefficient of ferroelectric nanoparticles under electric field, defined as the derivative of EC temperature change on external electric field,

$$\Sigma = \frac{d\Delta T_{EC}}{dE_{ext}}. \qquad (14)$$

The size effect of $\Sigma(E_{ext})$ can be important for ECE applications. Field dependence of $\Sigma$ calculated for several radii of BaTiO$_3$ nanoparticle and two frequencies are shown in **Figs. 9a-d**.

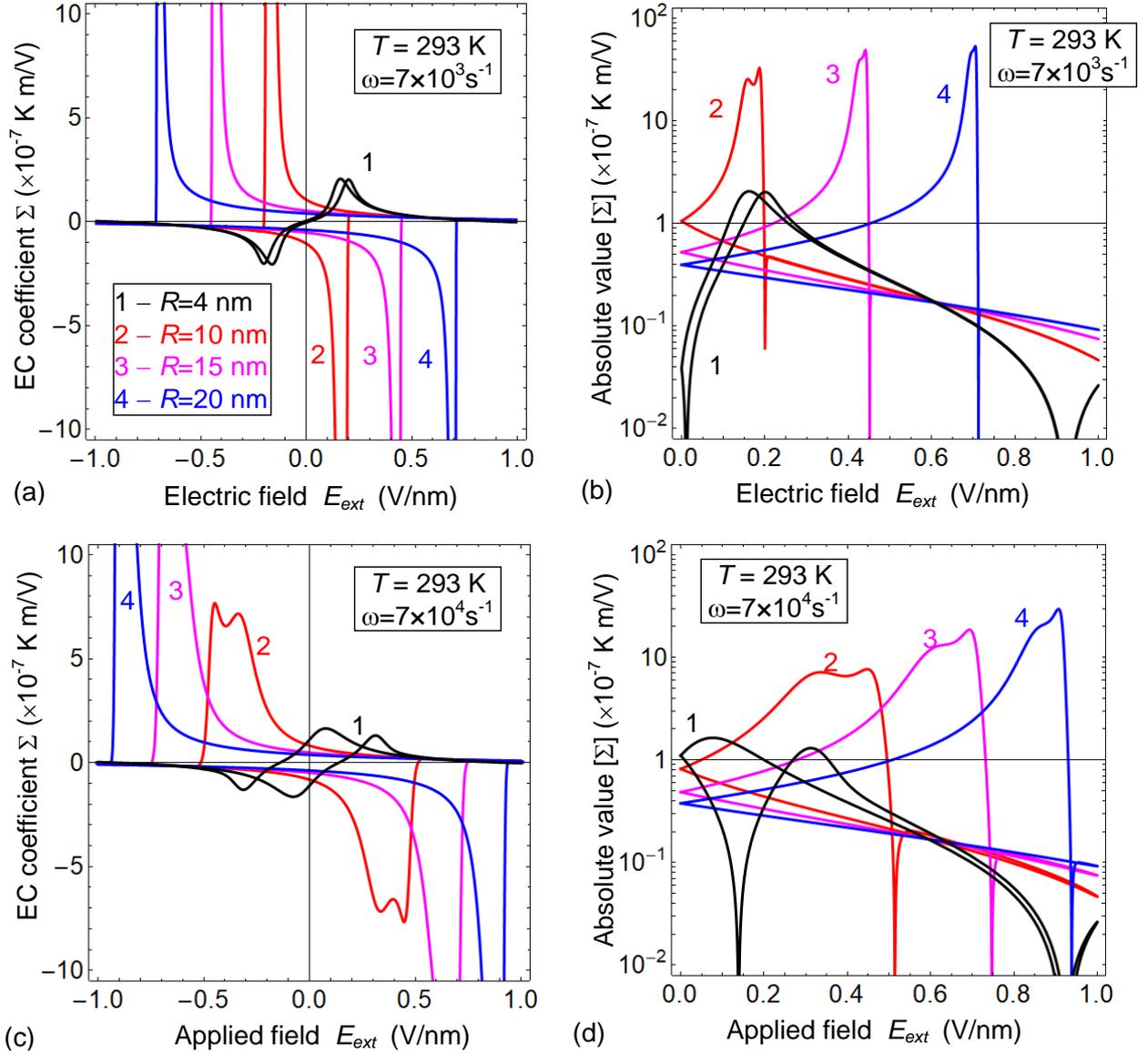

**FIGURE 9.** Dependences of EC coefficient $\Sigma$ **(a, c)** and its absolute value $|\Sigma|$ in logarithmic scale **(b, d)** on external electric field calculated for several radii $R = 4$, 10, 15, 20 nm of BaTiO$_3$ nanoparticle (curves 1-4), $T = 293$ K, frequency $\omega = 7 \times 10^3$ s$^{-1}$ **(a, b)** and $7 \times 10^4$ s$^{-1}$ **(c, d)**. Other parameters are the same as in **Fig. 3**.

In accordance with the field dependence $\Delta T_{EC}(E_{ext})$ shown in **Fig. 3d**, the dependence $\Sigma(E_{ext})$ is symmetrical with respect to the coordinate origin (**Fig.9a,c**). The field dependences $\Sigma(E_{ext})$ are characterized by the presence of a maximum, which position shifts to higher fields with increasing the nanoparticle size (see curves 1-4 in **Fig. 9b,d**). As we have discussed already, this maximum corresponds to the thermodynamic coercive field, which means the field induced entropy change is maximal at the polarization reversal in the single domain state. When the polarization switching occurs just via polarization rotation, the increasing entropy can be related to appearance of the polarization component perpendicular to the electric field direction. Similar effect has been observed in antiferroelectrics [85]. The appearance of a polydomain state in vicinity of the coercive field is quite probable in experiment, but it is not the case for a chosen small $\Lambda$ and high $\varepsilon_{IF}$.

Comparing the field dependences of $\Sigma$ shown in **Fig. 9c, d,** it is seen that the widths of maxima increase and their amplitudes decrease with increasing frequency. At higher frequencies, the dependence of the $\Sigma$ on $E_{ext}$ becomes more pronounced for a given size. At the same time in the unipolar range (along the upper or bottom branch of the hysteresis curve) the coefficient $\Sigma(E_{ext})$ continuously decreases with the increasing amplitude of the electric field. The decay of $\Sigma(E_{ext})$ is more pronounced for the larger particle size. The smallest paraelectric particle shows a maximum on the $\Sigma(E_{ext})$ dependence, which might be attributed to the critical point in a field-temperature phase diagram [86]. The smallest 4-nm particles, which size is well below the critical size (~8 nm at room temperature), are characterized by a zero value of $\Sigma$ at the zero field (curve 1), unlike the situation for nanoparticles of a larger diameter (curves 2-4), that corresponds to $\Delta T_{EC}(E_{ext})$ shown in **Fig. 3d**. Comparing the dependences $\Sigma(E_{ext})$ shown in **Fig. 9c, d,** it is seen that the widths of maxima increase and their amplitudes decrease with increasing frequency. At higher frequencies, the dependence of the $\Sigma$ on $E_{ext}$ becomes more pronounced for given particle size.

Analysis of the radius and electric field dependences of $\Sigma(E_{ext})$ presented in this subsection shows the possibility to control the effectiveness of EC energy conversion by changing the radius of a nanoparticle.

## VI. DISCUSSION AND CONCLUSION

Using LGD theory, we calculated and analyzed the dependences of polarization, dielectric permittivity, pyroelectric, and electrocaloric properties on external electric field, temperature, and radius of a spherical single-domain ferroelectric nanoparticle covered by a semiconducting shell and placed in a dielectric medium. The chosen geometry is typical for the theoretical consideration of ferroelectric nanocomposites in the effective medium approximation, if the volume fraction of the nanoparticles is relatively small (e.g. less than 10%).

For numerical simulations we considered BaTiO$_3$ nanoparticles placed in a polymer matrix, since such nanocomposites already exist and regarded attractive for pyroelectric and electrocaloric applications. We have chosen BaTiO$_3$ because it undergoes the first order phase transition from the ferroelectric to paraelectric phase, and this fact adds additional interesting peculiarities of PE and EC properties, such as the temperature and size-induced hysteresis, in comparison with the ferroelectric materials undergoing the second order phase transition. It should be noted that analytical expressions derived in the paper can be applied for any other ferroelectric material.

We established how the particle size determines the behavior PEE and ECE in the single-domain ferroelectric nanoparticles with the first order phase transition. We show that one can induce the maxima of PE coefficient and EC temperature variation, control their width, height, and sign by changing the particle size, as well as tune the voltage and temperature intervals for which PEE or/and ECE are maximal. Also we revealed that it is possible to select the interval of particle radii, for which PE and/or EC energy conversion are maximal at room temperature. Corresponding trends based on our calculations are summarized in **Table II**.

**Table II**. Size effect of polar, dielectric, PE and EC properties of single-domain ferroelectric nanoparticles with the 1$^{st}$ order FEPT

| Nanoparticle property | Size effect at small static external electric fields ($0 \leq E \ll E_{cr}$, $\omega=0$) | Influence of external quasi-static electric fields $E$ ($0 \ll E < 10 E_{cr}$, $\omega$ is zero or small[***]) |
|---|---|---|
| Ferroelectric polarization $P(T,R)$ [Figs.3a, 4a, 5a, 6a] | $P(T,R)$ disappears (at $E=0$) or becomes rather small (at $E>0$) for temperatures $T>T_{cr}(R)$[*] and sizes $R<R_{cr}(T)$[**]. Temperature and size hysteresis of $P(T,R)$ exists near $T \approx T_{cr}(R)$ and $R \approx R_{cr}(T)$, respectively. | External fields comparable with $E_{cr}$ induce irreversible polarization at $T>T_{cr}(R)$, the value of which increases with $E$ increasing. Increasing the field in the range $0 \ll E < 5 E_{cr}$ smooths out all temperature and size features of $P(T,R)$. The temperature hysteresis disappears at $E \geq E_{cr}$. $P(E)$ loops become wider and metastable states disappear with $\omega$ increase |
| Dielectric permittivity $\varepsilon_{NP}(T,R)$ [Figs.3b, 4b, 5b, 6b] | $\varepsilon_{NP}(T,R)$ is maximal (at $E>0$) or diverges (at $E=0$) at $T=T_{cr}(R)$ and $R=R_{cr}(T)$, respectively. Temperature and size hysteresis of $\varepsilon_{NP}(T,R)$ exists near $T \approx T_{cr}(R)$ and $R \approx R_{cr}(T)$, respectively. | Increasing the field in the range $0 \ll E < 5 E_{cr}$ significantly broadens $\varepsilon_{NP}(T)$ maximum, decreases the maximum height and shifts its $R$-dependent position to higher $T$. The temperature hysteresis disappears at $E \geq E_{cr}$. The maximum (attributed to the FEPT) disappears at $E \gg E_{cr}$. |
| PE coefficient, $\Pi(T,R)$ [Figs.3c, 4c, 5c, 6c] | $\Pi(T,R)$ is maximal (at $E>0$) or diverges (at $E=0$) at $T=T_{cr}(R)$ and $R=R_{cr}(T)$. $\Pi(T,R)$ almost vanishes at $T>T_{cr}(R)$ and $R<R_{cr}(T)$. Temperature and size hysteresis of $\Pi(T,R)$ exists near $T \approx T_{cr}(R)$ and $R \approx R_{cr}(T)$, respectively. | Increasing the field in the range $0 \ll E < 5 E_{cr}$ significantly broadens $\Pi(T)$ maximum, decreases the maximum height and shifts its $R$-dependent position to higher $T$. The temperature hysteresis disappears at $E \geq E_{cr}$. |
| PE detection FoM, $F_I(T,R)$ and $F_f(T,R)$ [Figs.7b-c, 8b-c] | FoM are maximal at $T=T_{cr}(R)$. The temperature hysteresis exists close to $R=R_{cr}(T)$. The hysteresis width increases with $R$ decrease. | Increasing the field in the range $0 \ll E < 5 E_{cr}$ significantly broadens $F_I(T)$ and $F_f(T)$ maxima and shifts their $R$-dependent position to higher $T$. The field increasing decreases the height of |

| | | $F_I(T)$ maximum, but does not change the height of $F_f(T)$ maximum. The temperature hysteresis of $F_I(T)$ and $F_f(T)$ disappears at $E \geq E_{cr}$. |
|---|---|---|
| PE energy conversion FoM $F_{EQ}(T,R)$ and $F_{EU}(T,R)$, and coupling factor $K_{PE}(T,R)$ [Figs.7d-f, 8d-f] | The factors are maximal at $T=T_{cr}(R)$. The temperature hysteresis exists close to $R=R_{cr}(T)$. The hysteresis width increases with $R$ decrease. | Increasing the field in the range $0<<E<5E_{cr}$ significantly broadens $F_{EQ}(T)$, $F_{EU}(T)$ and $K_{PE}(T)$ maxima, decreases their height and shifts the maxima position to higher $T$. The temperature hysteresis of PE energy conversion factors disappears at $E \geq E_{cr}$. |
| EC temperature change, $\Delta T_{EC}(T,R,E)$ [Figs.3d, 4d, 5d, 6d] | $\Delta T_{EC}(T,R)$ is zero at $E=0$. At $0<E<<E_{cr}$ $\Delta T_{EC}(T)$ is maximal at $T=T_{cr}(R)$. $\Delta T_{EC}(R)$ is positive below $R_{cr}(T)$ and changes its sign depending on the field magnitude at $R>R_{cr}(T)$. Temperature and size hysteresis of $\Delta T_{EC}(T,R,E)$ exists near $T \approx T_{cr}(R)$ and $R \approx R_{cr}(T)$, respectively. | Increasing the field in the range $0<<E<5E_{cr}$ significantly broadens $\Delta T_{EC}$ maximum, increases its height and shifts its $R$-dependent position to higher $T>T_{cr}(R)$. The temperature hysteresis disappears at $E \geq E_{cr}$. The frequency increase narrows the region of nonzero $\Delta T_{EC}$. |
| EC coefficient $\Sigma(T,R,E)$ [Fig.9] | $\Sigma(R)$ is close to zero at $R<R_{cr}(T)$, maximal at $R=R_{cr}(T)$ and decreases with $R$ increase at $R>R_{cr}(T)$. Narrowing of $\Sigma(E)$ hysteresis width and suppression of its maxima at $E=E_{coercive}$ occurs with $R$ decrease. | Increasing the field in the range $0<<E<5E_{cr}$ induces nonzero $\Sigma(E)$ at $R<R_{cr}(T)$. $\Sigma(E)$ hysteresis disappears at $E>>E_{cr}$. The coercive field of $\Sigma(E)$ hysteresis strongly increases with $R$ increase and then saturates to the bulk value. The frequency increase significantly broadens $\Sigma(E)$ maxima at coercive field, decreases peak values and shift them to higher fields. |

*$T_{cr}(R)$ is the critical temperature of the FEPT for a nanoparticle of radius $R$, which decreases with $R$ decrease, increases with $E$ increase, and can be estimated from Eq.(5b) for small fields $0 \leq E<<E_{cr}$. The FEPT at $T=T_{cr}(R)$ smears and eventually disappears at $E >> E_{cr}$.

**$R_{cr}(T)$ is the particle critical size at a temperature $T$, which decreases with $T$ and $E$ increase, and can be estimated from Eq.(11) for small fields $0 \leq E<<E_{cr}$. $R_{cr}(T)$ disappears at $E >> E_{cr}$.

*** The frequency effect is mentioned in the table only for those few cases, which are studied in the paper. The systematic study of $\omega$-effect will be performed elsewhere.

Allowing for the generality of performed consideration for ferroelectric nanoparticles with the first order phase transition of displacement type the obtained results are valid not only for BaTiO$_3$, but for many other ferroelectrics with the same type of phase transition (e.g. KNbO$_3$). It follows from **Table II** that PE coefficient, FoM, and energy conversion can be anomalously large for the particles with size near the critical one. The electrocaloric characteristics $\Delta T_{EC}$ and its derivative on external electric field have maximal value around the critical size, but they are not large, because the field mainly smears temperature dependence of polarization.

To summarize the obtained analytical results demonstrate possibilities to control the pyroelectric and electrocaloric properties of ferroelectric nanocomposites, and the working performances (figures of merits, energy conversion efficiency) of PE and EC convertors by changing the nanoparticle sizes, and tuning the amplitude and frequency of the external electric field. This is important for advanced cryogenic and energy harvesting applications. Using the ferroelectrics undergoing the first order phase transition adds additional interesting peculiarities of the PE and EC properties, such as the temperature and size-induced hysteresis, in comparison with ferroelectric materials undergoing the second order phase transition.

**Acknowledgements.** A.N.M. expresses her deepest gratitude to Prof. S.M. Ryabchenko (NASU), Prof. V.Yu. Reshetnyak (KNU) and Prof. R. Hertel (CNRS) for stimulating discussions about the nature of effective screening length, electrostatic problems and polarization vortexes. A.N.M. work was partially supported by the National Academy of Sciences of Ukraine (project No. 0118U003375 and No. 0117U002612) and by the Program of Fundamental Research of the Department of Physics and Astronomy of the National Academy of Sciences of Ukraine (project No. 0117U000240). M.V.S. acknowledges RFBR grant 17-08-01374_a and Russian academic excellence project "5–100″ for Sechenov First Moscow State Medical University. This project has received funding from the European Union's Horizon 2020 research and innovation programme under the Marie Skłodowska-Curie grant agreement No 778070.

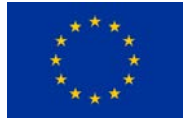

**Authors' contribution.** A.N.M. and V.V.S. generated research idea and proposed a model. A.N.M. stated the problem, derived analytical expressions jointly with E.A.E., interpreted results and wrote the manuscript draft. E.A.E. jointly with H.V.S. wrote the codes and performed numerical calculations. M.D.G., N.V.M., V.V.S., and other co-authors densely worked on the results discussion and manuscript improvement.

# APPENDIX A

## A1. Derivation of equation for polarization with renormalized coefficients

Let us consider the spherical ferroelectric particle with polarization **P** oriented along one of the principal crystallographic axis, denoted below as $z$. Here we also introduce an isotropic background permittivity $\varepsilon_b$ of ferroelectric particle. The media outside the particle is a dielectric with permittivity $\varepsilon_e$. Electrical displacement is $\mathbf{D}_i = \varepsilon_0\varepsilon_b\mathbf{E}_i + \mathbf{P}$ and $\mathbf{D}_e = \varepsilon_0\varepsilon_e\mathbf{E}_e$, where the subscript "*i*" means the physical quantity inside the particle, "*e*" – outside the particle; $\varepsilon_0$ is a universal dielectric constant. We introduce electric field $\mathbf{E} = -\nabla\varphi$ via electrostatic potential $\varphi$, which should satisfy Poisson and Laplace equations inside and outside the particle, respectively

$$\varepsilon_0\varepsilon_b\left(\frac{\partial^2}{\partial x^2} + \frac{\partial^2}{\partial y^2} + \frac{\partial^2}{\partial z^2}\right)\varphi_i = -\frac{\partial P}{\partial z} \quad \text{inside particle} \tag{A.1a}$$

$$\varepsilon_0\varepsilon_e\left(\frac{\partial^2}{\partial x^2} + \frac{\partial^2}{\partial y^2} + \frac{\partial^2}{\partial z^2}\right)\varphi_e = 0 \quad \text{outside particle} \tag{A.1b}$$

supplemented by the interface conditions of potential continuity at the particle surface S both for potential and normal components of electrical displacement:

$$\left(\varphi_e - \varphi_i\right)\big|_S = 0; \qquad \varphi_e\big|_{r\to\infty} = -z\,E_{ext}, \tag{A.2a}$$

$$\left(\left(\mathbf{D}_e - \mathbf{D}_i\right)\mathbf{n} + \varepsilon_0\varepsilon_{IF}\frac{\varphi_i}{\Lambda}\right)\bigg|_S = 0. \tag{A.2b}$$

Here **n** is the outer normal to the particle surface. In Eq. (A.2a) $E_{ext}$ is the external electric filed in the media far from the particle. In Eq.(A.2) we take into consideration the effective screening charge density, proportional to the surface potential and inversely proportional to effective screening length $\Lambda$. Below we suppose that the polarization gradient is small.

For a spherical particle the general solutions of Eqs.(A.1) could be expanded into the series on Legendre polynomials. For the considered problem a few terms are sufficient, namely

$$\varphi_e = E_e\frac{R^3}{r^2}\cos\theta - r\cos\theta\,E_{ext}, \qquad \varphi_i = -E_i r\cos\theta. \tag{A.3}$$

Here $\theta$ is the polar angle for spherical coordinate system, $r$ is the corresponding radial coordinate; $E_e$ and $E_i$ are the constants to be determined from the boundary conditions (A.2).

Substitution of Eqs.(A.2) to Eqs. (A.3) leads to condition

$$\left(E_e - E_{ext}\right)R\cos\theta = -E_i R\cos\theta \quad \Rightarrow \quad E_e = E_{ext} - E_i \tag{A.4}$$

Radial components of field could be obtained from (A.3) as follows

$$\left(\mathbf{E}_e\right)_r = 2E_e\frac{R^3}{r^3}\cos\theta + E_{ext}\cos\theta \qquad \left(\mathbf{E}_i\right)_r = E_i\cos\theta. \tag{A.5}$$

Corresponding displacement is

$$(\mathbf{D}_e)_r = \varepsilon_0 \varepsilon_e \left[ 2E_e \frac{R^3}{r^3} \cos\theta + E_{ext} \cos\theta \right], \qquad (\mathbf{D}_i)_r = \varepsilon_0 \varepsilon_b E_i \cos\theta + P\cos\theta \qquad (A.6)$$

Substitution of Eq.(A.2) to Eq.(A.3) and (A.6) leads to conditions

$$\varepsilon_0 \varepsilon_e (2E_e + E_{ext})\cos\theta - \varepsilon_0 \varepsilon_b E_i \cos\theta - P\cos\theta - \varepsilon_0 \varepsilon_{IF} \frac{E_i R \cos\theta}{\Lambda} = 0. \qquad (A.7a)$$

$$\Rightarrow \varepsilon_e (2E_e + E_{ext}) - \left(\varepsilon_b + \varepsilon_{IF} \frac{R}{\Lambda}\right) E_i = \frac{P}{\varepsilon_0}. \qquad (A.7b)$$

The solution of the linear system (A.4) and (A.7) has the form:

$$E_i = -\frac{P}{\varepsilon_0} \frac{1}{\varepsilon_b + 2\varepsilon_e + \varepsilon_{IF}(R/\Lambda)} + \frac{3\varepsilon_e}{\varepsilon_b + 2\varepsilon_e + \varepsilon_{IF}(R/\Lambda)} E_{ext}, \qquad (A.8b)$$

$$E_e = \frac{P}{\varepsilon_0} \frac{1}{\varepsilon_b + 2\varepsilon_e + \varepsilon_{IF}(R/\Lambda)}. \qquad (A.8b)$$

Below we use the expression (A.8) for the formulation of the phenomenological equations of state. Substituting electric field (A.8) into the LGD equation one obtains the equation for polarization in the sphere

$$\left(\alpha_T (T - T_C^*(R)) + \frac{1}{\varepsilon_0 (\varepsilon_b + 2\varepsilon_e + \varepsilon_{IF}(R/\Lambda))}\right) P + \beta P^3 = \frac{3\varepsilon_e}{\varepsilon_b + 2\varepsilon_e + \varepsilon_{IF}(R/\Lambda)} E_{ext} \qquad (A.9)$$

### A2. Free energy and material parameters

The compact form of the bulk polarization-dependent in a multiaxial ferroelectric with cubic parent phase is:

$$\Delta G_{FE} = a_i(T) P_i^2 + a_{ij} P_i^2 P_j^2 + a_{ijk} P_i^2 P_j^2 P_k^2 + g_{ijkl} \frac{\partial P_i}{\partial x_k} \frac{\partial P_j}{\partial x_l} - P_i E_i$$
$$- Q_{ijkl} \sigma_{ij} P_k P_l - s_{ijkl} \sigma_{ij} \sigma_{kl} - \frac{F_{ijkl}}{2} \left( \sigma_{ij} \frac{\partial P_k}{\partial x_l} - P_k \frac{\partial \sigma_{ij}}{\partial x_l} \right) \qquad (A.10)$$

Description and numerical values of the phenomenological coefficients $a_i$, $a_{ij}$, $a_{ijk}$ and gradient coefficients $g_{ij}$ included in Eq.(2b) can be found in **Table A1.** Electric field components $E_i$ are defined via electrostatic potential in the conventional way, $E_i = -\partial\varphi/\partial x_i$.

**Table A1.** Material parameters for bulk ferroelectric $BaTiO_3$

| coefficient | $BaTiO_3$ (collected and recalculated mainly from Ref. [a, b]) |
|---|---|
| **Symmetry** | Tetragonal at room temperature, m3m in a paraelectric phase |
| $\varepsilon_b$ | 7 (Ref. [b]) |

| $a_i$ (C$^{-2}\cdot$mJ) | $a_1$=3.34($T$−381)×10$^5$ (at 293°K −2.94×10$^7$) |
|---|---|
| $a_{ij}$ (C$^{-4}\cdot$m$^5$J) | $a_{11}$= 4.69($T$−393)×10$^6$−2.02×10$^8$, $a_{12}$= 3.230×10$^8$, (at 293°K $a_{11}$= −6.71×10$^8$ $a_{12}$= 3.23×10$^8$) |
| $a_{ijk}$ (C$^{-6}\cdot$m$^9$J) | $a_{111}$= −5.52($T$−393)×10$^7$+2.76×10$^9$, $a_{112}$=4.47×10$^9$, $a_{123}$=4.91×10$^9$ (at 293°K $a_{111}$= 82.8×10$^8$, $a_{112}$=44.7×10$^8$, $a_{123}$=49.1×10$^8$) |
| $Q_{ij}$ (C$^{-2}\cdot$m$^4$) | $Q_{11}$=0.11, $Q_{12}$= −0.043, $Q_{44}$=0.059 |
| $s_{ij}$ (×10$^{-12}$ Pa$^{-1}$) | $s_{11}$=8.3, $s_{12}$= −2.7, $s_{44}$=9.24 |
| $g_{ij}$ (×10$^{-10}$C$^{-2}$m$^3$J) | $g_{11}$=5.1, $g_{12}$= −0.2, $g_{44}$= 0.2 [c] |
| $F_{ij}$ (×10$^{-11}$C$^{-1}$m$^3$) | ~100 (estimated from measurements of Ref. [d]), $F_{11}$= +2.46, $F_{12}$=0.48, $F_{44}$=0.05 (recalculated from [e] using $F_{\alpha\gamma}=f_{\alpha\beta}s_{\beta\gamma}$) |

[a] A.J. Bell. Phenomenologically derived electric field-temperature phase diagrams and piezoelectric coefficients for single crystal barium titanate under fields along different axes. J. Appl. Phys. **89**, 3907 (2001).

[b] J. Hlinka and P. Márton, Phenomenological model of a 90° domain wall in BaTiO$_3$-type ferroelectrics. Phys. Rev. **B 74**, 104104 (2006).

[c] P. Marton, I. Rychetsky, and J. Hlinka. Domain walls of ferroelectric BaTiO$_3$ within the Ginzburg-Landau-Devonshire phenomenological model. Phys. Rev. **B 81**, 144125 (2010).

[d] W. Ma and L. E. Cross. Flexoelectricity of barium titanate. Appl. Phys. Lett. **88**, 232902 (2006).

[e] I. Ponomareva, A. K. Tagantsev, L. Bellaiche. Finite-temperature flexoelectricity in ferroelectric thin films from first principles. Phys.Rev. **B 85**, 104101 (2012)

## APPENDIX B. Heat capacity calculation details

$$\frac{\partial^2 g_R}{\partial T^2} \approx \frac{\partial}{\partial T}\left(\frac{\partial P}{\partial T}\left[\alpha_T(T-T_{cr})P + \beta P^3 + \gamma P^5 - \eta E_{ext}\right] + \frac{\alpha_T}{2}P^2 + \frac{\beta_T}{4}P^4 + \frac{\gamma_T}{6}P^6\right)$$
$$\equiv (\alpha_T P + \beta_T P^3 + \gamma_T P^5)\frac{\partial P}{\partial T} \equiv -\frac{(\alpha_T P + \beta_T P^3 + \gamma_T P^5)^2}{\alpha_T(T-T_{cr}) + 3\beta P^2 + 5\gamma P^4}$$
(B.1)

$$\left(C_P^0 + \frac{T(\alpha_T P + \beta_T P^3 + \gamma_T P^5)^2}{\alpha_T(T-T_{cr}) + 3\beta P^2 + 5\gamma P^4}\right)^{-1} \approx \frac{1}{C_P^0}\left(1 - \frac{T\overline{\chi}_E}{C_P^0}(\alpha_T P + \beta_T P^3 + \gamma_T P^5)^2\right)$$
(B.2)

Corresponding entropy change is given by expression, $\Delta S = -\int_{E_1}^{E_2}\left(\frac{\partial P}{\partial T}\right)_E dE$.

## APPENDIX C. Temperature dependences of the nanoparticle polar, PE and EC properties at high amplitudes of external field

The dependences $P(T)$, $\varepsilon_{NP}(T)$, $\Pi(T)$ and $\Delta T_{EC}(T)$ calculated for several nanoparticle radii, rather high amplitude of external field ($E_{ext}$ = 0.5 V/nm) and relatively low frequency $\omega$ = 2×10$^4$ s$^{-1}$ are shown in **Figs. C1a-d**. The temperature hysteresis, existing for small $E_{ext}$, disappears with $E_{ext}$ increasing, and the dependence *P(T)* significantly smears with $E_{ext}$ increasing. The effect is maximal for the smallest particle and decreases with the radius increase (compare solid curves 1-4 in **Fig. C1a**). The dependences *P(T)* in **Fig. C1a** are typical for the smearing of the 1-st order FEPT region by external field. Correlating with *P(T)* in **Fig. C1a**, $\varepsilon_{NP}(T)$ and $\Pi(T)$ in **Figs. C1b-c** show rather asymmetric maxima of the emerging at the phase transition temperature, which magnitude and sharpness noticeably increases with the particle radius increase, but the temperature position is almost radius-independent. **Figures C1b** and **C1c** look as if mirror symmetrical, but $\varepsilon_{NP}(T)$ and $\Pi(T)$ maxima displacement is opposite. This conclusion is valid for high amplitude of external electric field.

The dependences $\Delta T_{EC}(T)$ reveal several features shown in **Fig. C1d**. These features are distinctive from the ones shown in **Fig. 4d.** In particular, for the smallest particle (curve 1 in **Fig. C1d**) $\Delta T_{EC}$ value is nonzero and rather weakly changes [within (1.8 - 2.8 K)] in the wide temperature range. For bigger particles $\Delta T_{EC}$ value is also nonzero, but very small at T < 250 K, sharply increases and reaches ≈ 3 K in large vicinity of 350 K, and then gradually decreases at T > 350 K (see curves 2-4 in **Fig. C1d**). At that, the temperature range of the maximal EC conversion narrows and shifts to higher temperatures with the particle radius increase (compare the shape of maxima for the curves 2-4 in **Fig. 5d**), which is in correspondence with *P(T)* behavior (curves 2-4 in **Fig. C1a**) due to FEPT smearing and $T_{cr}$ shift.

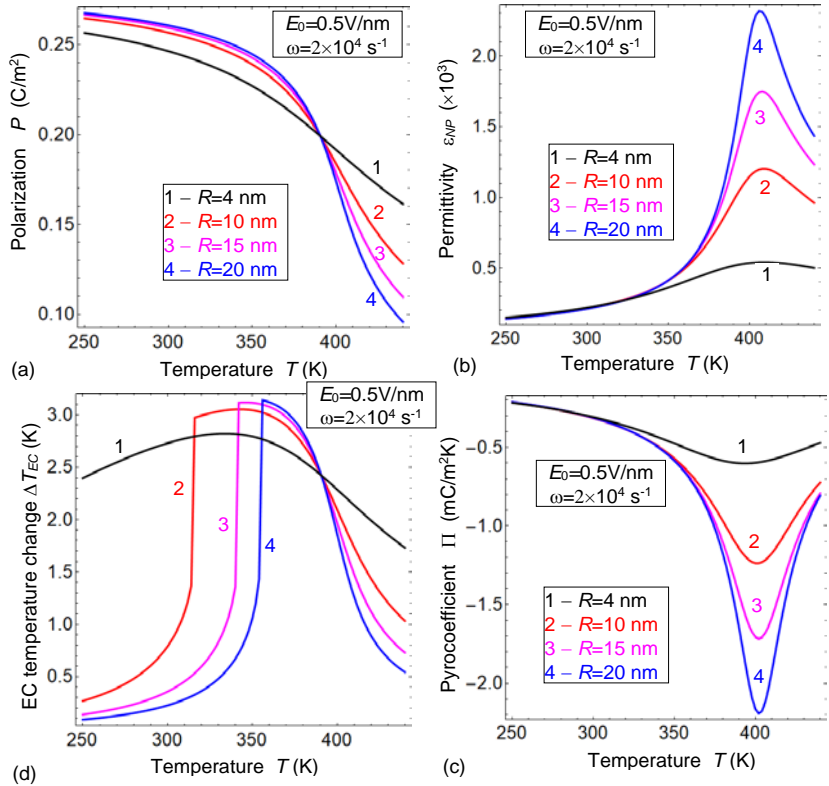

**FIGURE C1.** Temperature dependences of the BaTiO$_3$ nanoparticle polarization **(a),** relative dielectric permittivity **(b),** PE coefficient **(c)** and EC temperature change **(d)** calculated for several radii of BaTiO$_3$

nanoparticle $R$ = 4, 10, 15, 20 nm (curves 1-4). External field amplitude is 0.5 V/nm and frequency $\omega = 2\times10^4$ s$^{-1}$. Other parameters are the same as in **Fig. 3.**

As can be seen from the **Fig. C1c,d**, even at high fields we have the possibilities to control the temperature range and width of the maximal PE and EC conversion by changing the particle size.

## APPENDIX D. Pyroelectric figures of merits and other energy conversion factors

In the energy conversion operation mode, PE convertor of capacity $C$ generates pyro-charge $Q_\pi$ during thermal cycle with the same d$T$ value. At that, the electric energy generated during heating/cooling cycle is $W_\pi = 2(Q_\pi^2/2C) = (A\Pi dT)^2/\varepsilon\varepsilon_0 A/h = (\Pi^2/\varepsilon\varepsilon_0)Ah\cdot dT^2$. (Here $A$ is the electrode area, $h$ is the interdelectrode distance for flat capacitor.) For this case, the energy FoM in view $F_E = \Pi^2/\varepsilon\varepsilon_0$ have been proposed. If the PE convertor is subjected to an incident radiation (as pyroelectric detector [48, 95, 97, 98]), the electric energy generated during heating/cooling cycle is $W_\pi = 2(CU_\pi^2/2) = (\varepsilon\varepsilon_0 A/h)[(\Pi/\varepsilon\varepsilon_0)h\cdot dT]^2$. Substitution $dT = W_{th}/c_\rho Ah$ gives $W_\pi = (\varepsilon\varepsilon_0 A/h)[(\Pi/\varepsilon\varepsilon_0)h(W_{th}/c_\rho Ah)]^2 = (\Pi^2/c_\rho^2\varepsilon\varepsilon_0)(1/Ah)W_{th}^2$. For this case have been proposed energy FoM in view $F_E' = \Pi^2/c_\rho^2\varepsilon\varepsilon_0$.

The efficiency $\eta$ of the PE energy conversion is $\eta = W_\pi/W_{th}$, where $W_\pi = (\Pi^2/\varepsilon\varepsilon_0)Ah\cdot dT^2$ is the electric energy generated during heating/cooling cycle, and $W_{th} = c_\rho Ah\, dT$ is the heat energy, $A$ is the electrode area, $h$ is the inter-electrode distance for flat capacitor. Maximal efficiency $\eta_C$ given by ideal Carnot cycle is $\eta_C = dT/T_A$, where $T_A$ is the ambient temperature. Therefore, $\eta = W_\pi/W_{th} = (\Pi^2/c_\rho\varepsilon\varepsilon_0)dT = (\Pi^2/c_\rho\varepsilon\varepsilon_0)\eta_C T_A$ or $\eta = k^2\eta_C$, where $k^2$ is the pyroelectric (electro-thermal) coupling factor, $k^2 = (\Pi^2/c_\rho\varepsilon\varepsilon_0)\cdot T_A$.